\documentclass[lettersize,journal]{IEEEtran}
\usepackage{amsmath,amsfonts}
\usepackage{algorithmic}
\usepackage{algorithm}
\usepackage{booktabs}
\usepackage{array}
\usepackage{pdfpages}
\usepackage{subcaption}
\usepackage[inkscapearea=page]{svg}
\usepackage{listings}
\usepackage{xcolor}
\usepackage{enumitem}
\usepackage[font=small]{caption}

\definecolor{codegreen}{rgb}{0,0.6,0}
\definecolor{codegray}{rgb}{0.5,0.5,0.5}
\definecolor{codepurple}{rgb}{0.58,0,0.82}
\definecolor{backcolour}{rgb}{0.95,0.95,0.92}

\lstdefinestyle{mystyle}{
    backgroundcolor=\color{backcolour},   
    commentstyle=\color{codegreen},
    keywordstyle=\color{magenta},
    numberstyle=\tiny\color{codegray},
    stringstyle=\color{codepurple},
    basicstyle=\ttfamily\footnotesize, 
    breakatwhitespace=false,         
    breaklines=true,                 
    captionpos=b,                    
    keepspaces=true,                 
    numbers=left,                    
    numbersep=5pt,                  
    showspaces=false,                
    showstringspaces=false,
    showtabs=false,                  
    tabsize=2,
    frame=single,                    
    rulecolor=\color{black}
}
\usepackage[compact]{titlesec}
\usepackage[colorlinks=true, urlcolor=blue]{hyperref}
\usepackage{textcomp}
\usepackage{stfloats}
\usepackage{url}
\usepackage{verbatim}
\usepackage{graphicx}
\usepackage{cite}
\title{\texttt{bayesgrid}: An Open-Source Python Tool for Generating Probabilistic Synthetic Transmission-Distribution Grids Using Bayesian Hierarchical Models}

\author{Henrique O. Caetano, Rahul K. Gupta and Carlos D. Maciel

\thanks{Henrique O. Caetano is with the Department of Electrical and Computing Engineering, São Carlos School of Engineering, University of São Paulo, São Carlos,
SP, Brazil.}

\thanks{Rahul K. Gupta is with the School of Electrical Engineering \& Computer Science, Washington State University, USA}

\thanks{Carlos D. Maciel is with the Department of Electrical Engineering, School of Engineering and Science, State University of São Paulo}

\thanks{This work was partially fomented by São Paulo Research Foundation (FAPESP), grants 2021/12220-1, 2023/07634-7 and 2024/08485-8.}
}

\makeatletter
\patchcmd{\@maketitle}
  {\addvspace{0.5\baselineskip}\egroup}
  {\addvspace{-2\baselineskip}\egroup}
  {}
  {}
\makeatother
\begin{document}

\maketitle
\begin{abstract}
In this work, we present \texttt{bayesgrid}, an open-source python toolbox for generating synthetic power transmission-distribution systems for any geographical location worldwide, using the publicly available data from OpenStreetMap (OSM). The toolbox is based on Bayesian Hierarchical Models (BHM) which is trained on existing distribution network databases to develop a probabilistic model and can be applied to any geographical location worldwide, leveraging transfer learning. Thanks to the BHM, the tool is capable of generating multiple instances of the distribution system for a same region. The generated networks contain three-phase phase-consistent unbalanced networks, radial topology and information on the nodal demand distributions. The generated network also contain the critical reliability indices, specifically the interruption duration and frequency of failure for individual grid components, allowing its application in reliability-related studies. The tool is demonstrated for different case studies generating synthetic network datasets for different geographical regions around the world. The framework allows saving the generated networks into open-source platforms: \textit{Pandapower} and OpenDSS. We also present an application for computation of probabilistic hosting capacity using the synthetic networks.
\end{abstract}
\section{Introduction}
\subsection{Motivation}
The robust planning, analysis, and simulation of electric power distribution systems fundamentally depend on access to detailed, granular network data, including topology, component parameters, and demand profiles \cite{li2022review}. However, real-world utility data is frequently inaccessible to the broader research community due to strict confidentiality policies, security concerns, and their commercially sensitive nature 
\cite{li2020building, Caetano2024}. This scarcity creates a significant barrier to innovation, making it difficult for researchers to validate new algorithms, benchmark optimization methods, or assess the scalability of proposed solutions for modern distribution system planning and control.

To overcome this challenge, the generation of synthetic power networks has emerged as an essential alternative \cite{mateo2018european,schweitzer2018mathematical,amme2018ego,sadeghian2020autosyngrid,li2020building,mateo2020building, meyur2022ensembles, banze2024open}. Synthetic systems provide versatile datasets that mimic the statistical and physical characteristics of real-world networks without disclosing sensitive information \cite{Ali2023, mateo2020building}. For power system applications, these synthetic networks must preserve electrical realism, such as plausible impedances, phase connectivity, and voltage behavior, to be useful for state estimation, optimal power flow, and reliability analysis \cite{Caetano2025_powertech}. Furthermore, as the grid transitions toward more active and decentralized operation, there is a growing need to model complex characteristics often overlooked in transmission-level synthesis, such as unbalanced three-phase flows and the stochastic nature of reliability indices \cite{Fogliatto2022}.

Despite the clear need for such data, the availability of open-source tools to generate these networks remains limited. Many existing methodologies rely on rule-based heuristics or closed datasets which makes it non-reproducible. Also, they produce deterministic outputs that fail to capture the inherent uncertainty of distribution systems. An open-source, probabilistic framework is therefore critical for democratizing access to realistic test cases. By providing a tool that can learn from real data and generate ensembles of valid networks, researchers and operators can conduct statistical analyses of grid performance under a wide range of topological and operational scenarios.

\subsection{Literature Review of Existing Tools}
Several recent works have developed open-source tools for generating synthetic power grids to address the lack of public test systems. These tools largely rely on open data sources, 
yet they often differ significantly in their modeling scope, flexibility, and electrical details.

In \cite{pylovo_ref_2025}, the authors present \textit{pylovo} (Python tool for Low-Voltage distribution grid generation). This tool utilizes a graph-based approach and data clustering to generate scalable grid solutions for large areas, such as cities or states. It incorporates detailed infrastructure planning, including greenfield and brownfield transformer positioning and multiple feeders. However, \textit{pylovo} relies heavily on specific building data, which can be difficult to generalize across different regions, and does not support user-provided utility data for custom calibration. Furthermore, it generates a single deterministic network instance and does not account for reliability indices or unbalanced three-phase power flows.

Similarly, the \textit{REGAL} (Reference Electricity Grid Analysis) model \cite{regal_ref_2024} offers a methodology to create synthetic low-voltage (LV) grids for country-sized areas based on open data. While effective for macro-scale analysis, it employs strong simplifications, such as defining grids within fixed cells and assuming a uniform distribution of consumers. Like \textit{pylovo}, it produces a single network instance and lacks the capability to model three-phase unbalance or reliability metrics.

Targeting a specific geographic context, the work in \cite{shift_ref_2024} introduces a tool that utilizes building data from OpenStreetMap to generate grids for locations in India. While it successfully integrates open data to produce OpenDSS-compatible files, these networks are highly specific to the Indian context and require defining multiple technical input parameters. It does not support probabilistic generation of multiple grid instances, nor does it model reliability or detailed phase unbalance.

Focusing on the topological aspects of generation, the authors in \cite{Giacomarra2024} propose a method for generating synthetic transmission grids using Exponential Random Graph (ERG) models. Their approach rigorously captures topological nuances, such as edge counts and triangle formations, ensuring the generation of connected graphs. However, this method is purely topological; it does not learn electrical parameters from utility data, does not integrate with OpenStreetMap for geographic realism, and neglects both reliability indices and unbalanced three-phase modeling.

In a recent contribution, the authors in \cite{weber2024open} introduced an automated, open data-driven methodology for modeling residential distribution grids. Their tool utilizes 2D and 3D building footprint data from OpenStreetMap to estimate consumer loads and employs deterministic optimization algorithms to geographically route 20 kV and 400 V grid topologies. While this approach is highly effective for structural planning and load estimation using minimal data, it is deterministic, producing a single static grid layout for a given set of inputs. Consequently, it lacks the probabilistic capabilities required to generate multiple stochastic instances for uncertainty quantification. Furthermore, the methodology assumes balanced grid conditions, neglecting the complexities of three-phase unbalance, and does not incorporate reliability indices.

Apart from these open-source tools, the generation of synthetic grids has gained increasing attention in the broader literature, with works such as \cite{li2020building, Ali2023, mateo2020building} proposing various methodologies. However, these studies generally do not provide an open-source software implementation or rely on closed datasets, limiting their reproducibility and widespread adoption.

\subsection{Main Contributions}
This paper introduces \texttt{bayesgrid}\footnote{Available at \url{https://github.com/HenriqueCaetano1/bayesgrid}}, a Python probabilistic toolbox designed to address the scarcity of realistic test systems for modern power systems research. While the validation of the underlying methodology for generating synthetic networks was established in our previous works \cite{caetano2026bayesian, Caetano2025_powertech,Caetano2024}, this paper focuses specifically on the open-source implementation of the tool, as well as user usage and the extensive of its application to any geographical region.

The primary objective of \texttt{bayesgrid} is to generate synthetic unbalanced three-phase distribution grids that explicitly incorporate reliability indices. To ensure topological realism, the framework integrates with OpenStreetMap (OSM) to derive network structures from actual street data, enabling the creation of georeferenced synthetic grids for virtually any location worldwide. Unlike deterministic approaches that output a single static network, \texttt{bayesgrid} leverages the stochastic nature of Bayesian Hierarchical Models (BHM) to quantify uncertainty and generate ensembles of possible grid realizations.
The key contributions of the proposed framework are summarized below and distinguished against the existing works in Table~\ref{tab:tool_comparison}.
\begin{itemize}[leftmargin=*]
    \item The tool introduces the capability to generate synthetic and probabilistic three-phase unbalanced distribution systems. Aside from usual electrical parameters, i.e., three-phase power demand and line parameters, it also incorporates critical reliability indices, specifically the duration and frequency of failure for individual grid components, enabling its application in large-scale reliability studies.  The tool can generate joint transmission-distribution systems, allowing joint coordination as well as planning studies, such as, hosting capacity optimization, etc. 
    \item The framework is integrated with OSM database, enabling the generation of synthetic grids for any location worldwide. Although trained on a massive dataset from a national Brazilian utility, the model is designed to accept custom user data, ensuring adaptability and broad applicability for both research and industrial purposes.
    \item The tool utilizes stochastic BHM framework allowing to quantify uncertainties inherent in grid synthesis. Given that synthetic generation is primarily employed when real-world data is unavailable, the reliance on learned data patterns makes uncertainty an unavoidable aspect of the process. To address this, the tool generates a collection of possible grid realizations rather than a single deterministic instance, thereby accounting for the uncertainty present in these predictive models.
\end{itemize}

\begin{table}[htbp]
\vspace{-1em}
\centering
\caption{Comparison of Synthetic Grid Generation Tools.}
\label{tab:tool_comparison}
\begin{tabular*}{\columnwidth}{@{\extracolsep{\fill}}lcccccccc}
\toprule
\textbf{Tool/Ref$^*$.} & \textbf{A1} & \textbf{A2} & \textbf{A3} & \textbf{A4} & \textbf{A5} & \textbf{A6} & \textbf{A7} & \textbf{A8} \\
\midrule
eGo-grid \cite{amme2018ego} & \checkmark & \checkmark & - & - & - & - & \checkmark & \checkmark \\
\cite{distribuicoes_parametros_sd_2017} & - & - & - & \checkmark & - & - & - & - \\
\textit{pylovo} \cite{pylovo_ref_2025} & \checkmark & \checkmark & - & - & - & -  & - & \checkmark \\
\textit{REGAL} \cite{regal_ref_2024} & \checkmark & \checkmark & - & - & - & - & - & - \\
\textit{SHIFT} \cite{shift_ref_2024} & \checkmark & \checkmark & - & - & - & - & \checkmark & - \\
\textit{ERG} \cite{Giacomarra2024} & \checkmark & \checkmark & - & \checkmark & - & - & - & - \\
\cite{Ali2023} & \checkmark & \checkmark & - & - & - & - & - & \checkmark \\
\cite{li2020building} & - & \checkmark & - & - & - & - & \checkmark & \checkmark \\
\cite{mateo2020building} & - & \checkmark & - & - & - & - & - & \checkmark \\
DAVE \cite{banze2024open} & \checkmark & \checkmark & - & - & - & - & \checkmark & \checkmark \\
\cite{meyur2022ensembles} & - & \checkmark & - & \checkmark & - & - & - & \checkmark \\
\textit{AutoSynGrid} \cite{sadeghian2020autosyngrid} & \checkmark & - & - & \checkmark & - & - & - & - \\
\cite{schweitzer2018mathematical} & - & \checkmark & - & \checkmark & - & - & - & - \\
\cite{weber2024open} & \checkmark & \checkmark & - & - & - & - & - & \checkmark \\
\textbf{\texttt{bayesgrid}} & \textbf{\checkmark} & \textbf{\checkmark} & \textbf{\checkmark} & \textbf{\checkmark} & \textbf{\checkmark} & 
\textbf{\checkmark} & \textbf{\checkmark} & 
\textbf{\checkmark} \\
\bottomrule
\end{tabular*}
\footnotesize\textsuperscript{$*$} \textbf{A1}: Open Source Software; \textbf{A2}: Uses Open Data; \textbf{A3}: Trainable on User-defined Datasets for grid parameters, power demand and failure events; \textbf{A4}: Multiple Stochastic Instances for the generated grid; \textbf{A5}: 3-Phase Consistent Unbalanced Modeling; \textbf{A6}: Consider Reliability Indices; \textbf{A7}: Joint transmission-distribution dataset; \textbf{A8}: Georeferenced grid topology
\vspace{-1em}
\end{table}
%
\section{Presenting \texttt{bayesgrid}}
\begin{figure*}[!htbp]
    \centering
    \includegraphics[width=\linewidth]{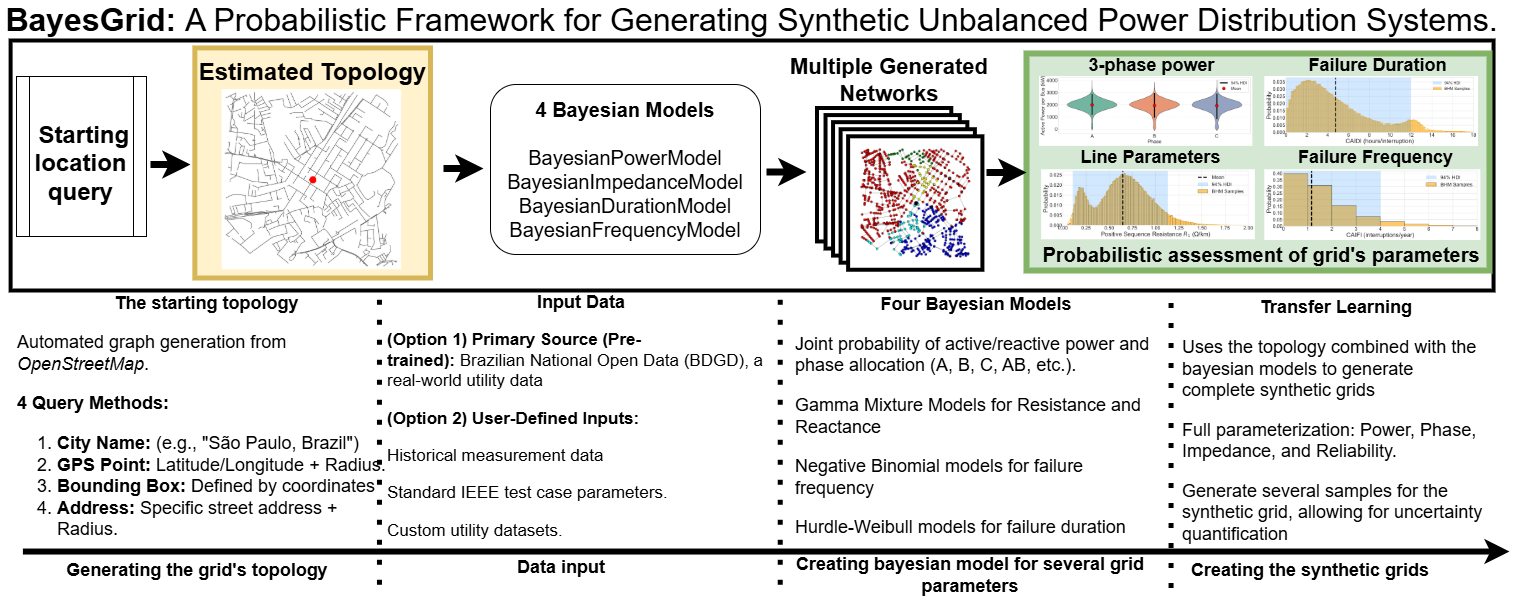}
    \vspace{0.1em}
    \caption{Overview of the \texttt{bayesgrid} framework. The workflow proceeds from data acquisition to the training of Bayesian models, the generation of network topology via OpenStreetMap, and finally the synthesis of grid ensembles through transfer learning.}
    \label{fig:bayes_grid_overview}
    \vspace{-1em}
\end{figure*}

This section presents the \texttt{bayesgrid} tool and description of the algorithms behind it. A detailed technical description of these algorithms can be found in author's previous works in \cite{Caetano2025_powertech, caetano2026bayesian}.
Figure \ref{fig:bayes_grid_overview} presents an overview of the framework, which consists of four main pillars: input standardization, topology estimation, Bayesian modeling, and probabilistic network generation. As it is shown, we start with OSM data to estimate the network topology and in parallel we train different BHM models using existing network database. Using the BHM models and topology, different instance of the synthetic networks are generated. In the following, we describe each step as well as the input data.
\subsection{Input Data}
\label{subsec:input_data}
The \texttt{bayesgrid} tool standardizes the required input data and supports various datatypes: \texttt{.csv}, \texttt{pandas} DataFrames, or Python lists. To train the BHM for synthetic networks using custom data, the input data must satisfy two primary requirements: a \textit{distance metric} and \textit{specific feature columns}.

The \textit{distance metric} refers to the distances between the nodes within the network and is used 
for generating distance-aware synthetic grids, which uses the \textit{hop distance}, an integer value ranging from 0 (substation) to a user-defined maximum. 
Users may provide a custom pre-calculated distance metric in the input dataset. Alternatively, if this metric is unavailable, the framework includes internal distance learning functions to compute it automatically. Users can select from topological-based, electrical-distance-based, or Euclidean-distance-based methodologies to generate the required discrete spatial layers for the model. 

The \textit{specific feature columns} refers to domain-specific information for different modules, and user must provide a DataFrame containing these information, which are
\begin{itemize}[leftmargin=*]
    \item \textbf{Reliability Frequency:} A column representing the count of interruptions per year (e.g., integers) of each bus.
    \item \textbf{Reliability Duration:} A list or column of failure durations (in hours) of each bus.
    \item \textbf{Power and Phase:} The per-phase power magnitudes at every bus, alongside the phase configuration (e.g., A, B, C, ABC).
    \item \textbf{Line Parameters:} The positive sequence resistance ($R$) and reactance ($X$) for each distribution line.
\end{itemize}

\subsection{Topology Estimation}
Before statistical parameters can be applied, the grid's topology needs to be defined. \texttt{bayesgrid} addresses the challenge of creating realistic grid topologies using the OSM data. 
First, the user defines the region of interest (e.g., city name, GPS coordinates, or bounding box). Then, the tool queries OSM to identify all available substations\footnote{\url{https://wiki.openstreetmap.org/wiki/Tag:power\%3Dsubstation}} and the underlying transportation network within that area. In scenarios where multiple substations are detected, the framework must determine which customers are supplied by which substation. To resolve this, a Voronoi-based algorithm is employed to partition the region into distinct service areas, assigning customers to the nearest substation geographically, a method motivated by approaches in previous literature \cite{Gupta2021, Mateo2024}.

Once the service areas are defined, the distribution topology is constructed based on the street map found in OSM. To ensure the grid operates radially, a standard requirement for distribution systems, a spanning tree algorithm is applied for each substation. This approach, motivated by our previous works \cite{caetano2026bayesian,Caetano2025_powertech,Caetano2024}, roots the tree at the substation and extends through the road network to reach all customers.

Following the definition of the feeder topology, distribution-level transformers are connected using a topology and clustering-based approach \cite{Mateo2020}. The transformers parameters are taken from \cite{Birchfield2017}. The framework allows for customizable voltage levels; for the results presented in this work, a medium voltage level of 13.8 kV and a low voltage level of 220 V are used, adhering to Brazilian standards \cite{bdgd_ref}.

Finally, the framework integrates the \textit{transmission and generation levels} to complete the grid hierarchy. Generators and transmission lines are extracted directly from OSM\footnote{\url{https://wiki.openstreetmap.org/wiki/Tag:power\%3Dline}} to form the upper-level topology. Since specific electrical parameters for these transmission components are often missing in open data, they are populated using standard values and transmission-level transformer data from the literature \cite{Birchfield2017}.

It is worth noting that the problem of estimating power system topology (for both transmission and distribution levels) has been extensively discussed and validated in the literature \cite{Taylor2024, Mateo2024}; therefore, the specific details and the validation of topological correctness is not presented in this work. Here, we focus on the probabilistic modeling of the operational parameters of the system, as it will be detailed next. 
\subsection{BHM Models}
At the core of the synthetic network generation engine is the BHM, a statistical approach built on several key pillars \cite{mcglothlin2018bayesian}.  \textit{First}, every unknown quantity, from line impedance to phase allocation, is treated as a random variable. Initial beliefs are specified as \textit{prior distributions}, which are updated via Bayesian inference to yield \textit{posterior distributions} once the model is fitted to observed data. \textit{Second}, this framework allows for the direct quantification of uncertainty by analyzing these posterior distributions. \textit{Third}, the model is hierarchical, allowing the distribution of variables (e.g., failure frequency) to depend on latent parameters (e.g., topological distance). \textit{Fourth}, the BHM enables efficient sampling from learned posterior distributions, allowing for the rapid generation of synthetic networks where each instance represents a statistically valid sample drawn from the model's learned distributions. 

The proposed framework instantiates this BHM approach through four Bayesian models, each targeting a specific aspect of the synthetic grid:
\begin{enumerate}[leftmargin=*]
    \item \textbf{\texttt{BayesianPowerModel}:} Captures the joint distribution of the phase configuration (e.g., single-phase versus three-phase) and the active power demand, ensuring that the generated phase allocations are topologically consistent with the network hierarchy.
    \item \textbf{\texttt{BayesianImpedanceModel}:} Characterizes the physical properties of network branches by learning distributions of positive-sequence resistance and reactance that reflect the varying conductor types at different distances from the substation.
    \item \textbf{\texttt{BayesianFrequencyModel}:} Estimates the occurrence rate of interruption events (CAIFI) using count-based distributions.
    \item \textbf{\texttt{BayesianDurationModel}:} Predicts the restoration time for those events (CAIDI) using survival analysis techniques.
\end{enumerate}

All four models are hierarchically conditioned on the grid's topology parameters, allowing the generated parameters to reflect the spatial heterogeneity of typical real-world distribution feeders.
\subsection{Probabilistic Network Generation}
Using the estimated topology and the learned BHM models, \texttt{bayesgrid} can be used to generate ensembles of synthetic networks. 
This process leverages \textit{Transfer Learning} to connect the training data with the target area (e.g., an OSM graph).
This network generation includes assigning active/reactive power, phase allocation (A, B, C, AB, etc.), line resistance and reactance, and reliability metrics (CAIFI and CAIDI) to specific nodes and edges. Since these parameters are sampled from the posterior distributions trained in the previous step, the tool outputs a set of samples rather than a single deterministic result. This approach enables robust Monte Carlo simulations and uncertainty quantification in downstream analysis.
\section{Main Modules and User Usage}
This section details the operational structure of \texttt{bayesgrid} tool, describing the primary modules available to the user. While the framework's core methodology relies on BHM, the implementation abstracts this complexity into high-level classes. For an extensive technical description, statistical validation of these models and their predictive and learning performance, the reader is referred to previous works \cite{caetano2026bayesian,Caetano2025_powertech,Caetano2024}.

\begin{figure*}[!htbp]
    \centering
    \includegraphics[width=0.9\linewidth]{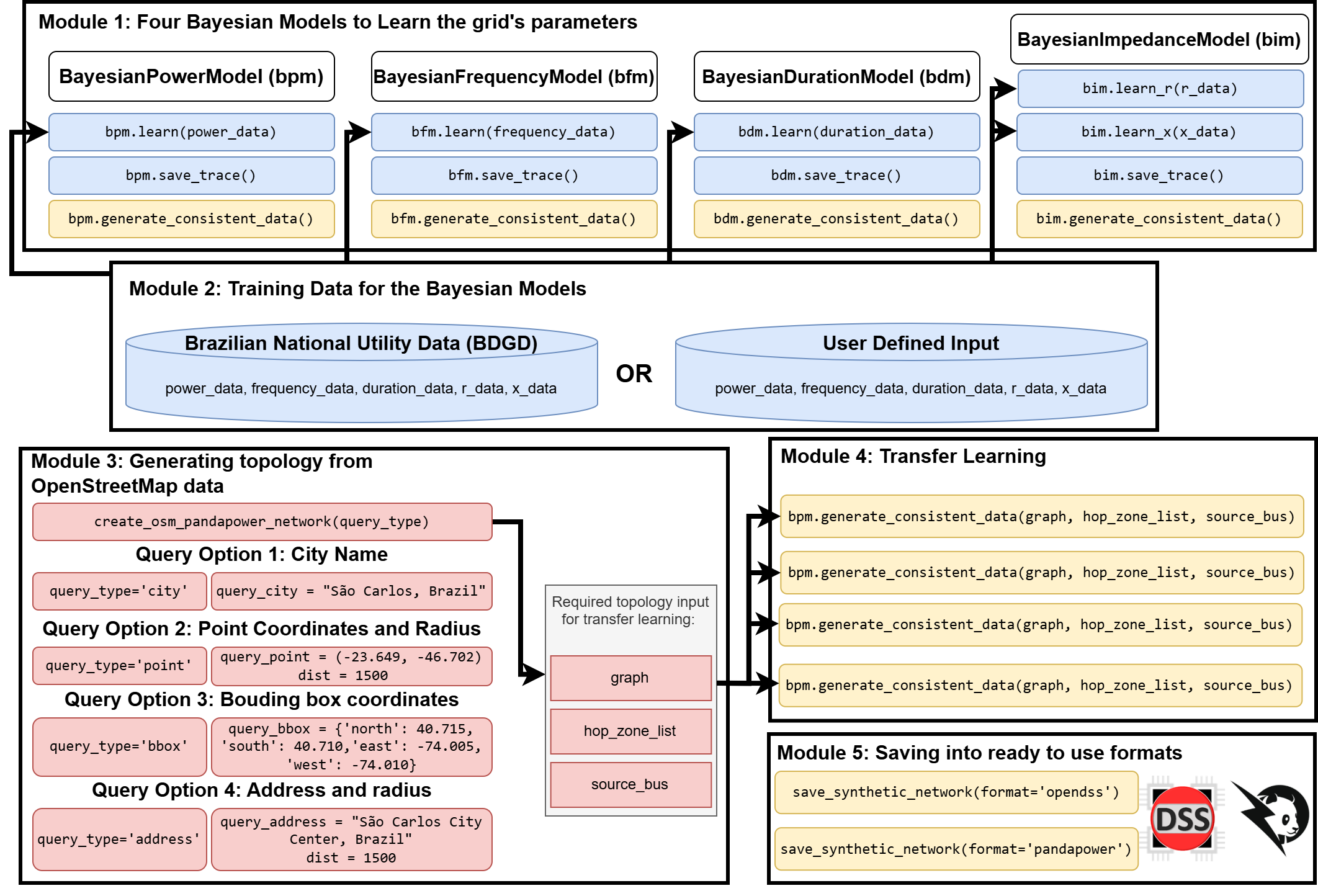}
    \caption{Overview of the main modules presented in \texttt{bayesgrid} tool.}
    \label{fig:bayes_grid_main_modules}
    \vspace{-1.0em}
\end{figure*}

The tool is organized into five primary modules, as illustrated in Figure~\ref{fig:bayes_grid_main_modules}: 1) The core Bayesian learning engines; 2) Custom model training capabilities; 3) Topology generation via OpenStreetMap; 4) Transfer learning for grid synthesis; and 5) Converting the generated synthetic utilities into practical formats for standard power simulation platforms.
\subsection{Module 1: Core Bayesian Models}
The probabilistic engine of \texttt{bayesgrid} is encapsulated in four distinct classes. Each class is responsible for learning and generating specific attributes of the distribution system. By default, these models are pre-trained on a massive dataset from the Brazilian National Utility Data (BDGD) \cite{bdgd_ref}, allowing users to generate realistic grids without providing their own training data.

The primary topological predictor used throughout the model is the concept of the \textit{discrete hop distance zone}, which serves as the primary latent variable connecting topology to electrical characteristics. For any bus $v$ in a graph $G=(V,E)$, we first calculate its continuous path distance (in km) from the substation. To capture non-linear dependencies (such as the tendency for load density or conductor types to change as one moves further from the feeder head) this distance is quantized into $Z$ distinct zones, denoted as $z_v \in \{1, \dots, Z\}$. Similarly, a line segment $l$ is assigned a zone $z_l$ based on its upstream node. This abstraction is fundamental to the framework's transfer learning capability: by conditioning all probabilities on a normalized topological metric ($z$) rather than absolute coordinates, the Bayesian models learn generalized spatial patterns that can be directly mapped onto any new network topology, regardless of its specific geographic layout. Each Bayesian model is described below.
\begin{enumerate}[leftmargin=*]
    \item \texttt{BayesianPowerModel}: This module learns the joint probability distribution of active power demand and phase allocation. To ensure physical realism, it enforces phase consistency (downstream phases must be a subset of upstream phases) using a topological scanning algorithm validated in \cite{Caetano2025_powertech}. Building data is incorporated as a normalization parameter in the prior distributions to account for load magnitude variations, as established in \cite{caetano2026bayesian}.

    The power demand $\mathbf{P}_v$ for a bus $v$ is conditioned on its phase configuration $\phi_v$ and its topological distance zone $z_v$. The model uses a hierarchical structure where the mean power $\boldsymbol{\mu}_{\phi_v}$ is derived from latent potential power distributions ($P^{\text{pot}}$) specific to single-, two-, and three-phase loads:
    \begin{align}
        P_v \mid \phi_v, z_v \sim \texttt{TruncatedNormal}(\boldsymbol{\mu}_{\phi_v}, \sigma_p^2, \text{lower}=0)
    \end{align}
    Phase allocation probabilities are learned via a Dirichlet-Categorical model conditioned on the zone $z_v$:
    \begin{small}
        \begin{align}
            \mathbf{c}_v \mid z_v \sim \texttt{Dirichlet}(\mathbf{A}_{z_v}); \quad \phi_v \sim \texttt{Categorical}(\mathbf{c}_v)
        \end{align}
    \end{small}
    \item {\texttt{BayesianImpedanceModel}}
This module determines the physical line parameters. It learns the positive-sequence resistance ($R_{l,1}$) and reactance ($X_{l,1}$) for line segments. To capture the multimodal nature of conductor types (e.g., different gauges for main feeders vs. laterals), the module employs a Gamma Mixture Model with $K=3$ components. The mixture weights $\mathbf{w}_{z_l}$ depend on the line's topological zone $z_l$:
\begin{align}
    R_{l,1} \mid z_l \sim \sum_{k=1}^{K} w^R_{z_l, k} \cdot \texttt{Gamma}(\alpha^R_k, \beta^R_k)
\end{align}
A similar structure is used for the $X/R$ ratio, ensuring realistic impedance characteristics.
\item {\texttt{BayesianFrequencyModel}}
To model reliability, specifically the Customer Average Interruption Frequency Index (CAIFI), this module uses a Negative Binomial distribution. This choice accounts for the overdispersed nature of failure count data $C_v$:
\begin{align}
    C_v \mid z_{v} \sim \texttt{NegativeBinomial}(\mu = \mu^{f}_{z_v}, \alpha = \alpha_{\text{dispersion}})
\end{align}
where the mean failure rate $\mu^{f}_{z_v}$ is hierarchical and varies by zone.
\item {\texttt{BayesianDurationModel}}
This module models the Customer Average Interruption Duration Index (CAIDI) using a two-part Hurdle-Weibull model. It first predicts the probability of a bus having \textit{any} interruption ($\lambda_v=1$) using a Bernoulli process, and then models the duration $D_v$ for those buses using a Weibull distribution:
\begin{align}
    \lambda_v \mid z_v &\sim \texttt{Bernoulli}(p_{z_v}) \\
    D_v \mid \lambda_v=1, z_v &\sim \texttt{Weibull}(\alpha_{z_v}, \beta_{z_v})
\end{align}
\end{enumerate}
\subsection{Module 2: Learning from data}
While the pre-trained BHM models can be used intermediately, the \texttt{bayesgrid} tool can also be re-trained using user-defined datasets. 
(e.g., utility measurements or standard IEEE test cases) using the \texttt{.learn()} method. This process utilizes the No-U-Turn Sampler (NUTS) to infer new posterior distributions, effectively re-calibrating the tool to reflect the specific statistical characteristics of the user's local grid environment. More details about the accepted formats for user data is given in Section \ref{subsec:input_data}.
\subsection{Module 3: Topology Generation}
To apply the Bayesian models to a new region, a physical grid topology is required. This module integrates with \textit{OSMnx} \cite{ref_osmnx_2017} to query OpenStreetMap \cite{OpenStreetMap} and generate realistic grid topologies. Users can generate a graph topology using a single function, \texttt{create\_osm\_pandapower\_network}, which accepts a \texttt{query\_type} argument to define one of four distinct retrieval methods:
\begin{enumerate}
    \item \texttt{query\_type=}=``city": Retrieves the street network for a named administrative boundary provided in the \texttt{query\_city} argument (e.g., ``São Carlos, Brazil").
    \item \texttt{query\_type=}``point": Captures the network within a specified radius \texttt{dist} from a central GPS coordinate tuple provided in \texttt{query\_point}.
    \item \texttt{query\_type=}``bbox": Defines a custom rectangular area via a dictionary of coordinates passed to \texttt{query\_bbox}.
    \item \texttt{query\_type=}``address": Geocodes a specific string provided in \texttt{query\_address} and retrieves the surrounding network within the radius defined by \texttt{dist}.
\end{enumerate}
This module automatically computes the specific topological metrics required as inputs for the transfer learning process in Module 4. As shown in the workflow (Module 3 in Figure \ref{fig:bayes_grid_main_modules}), the function returns the processed graph, identifies the specific index of the substation node ($v_s$) based on OSM tags, and calculates the discrete hop distance zone for every bus ($z_v$) and line ($z_l$) in the new system.
\subsection{Module 4: Transfer Learning and Synthesis}

To generate the synthetic data, the user invokes specific generation methods for each one of the four bayesian models, using the \texttt{.generate\_consistent\_data()} function, passing the new \texttt{graph}, the \texttt{hop\_zone\_list}, and the \texttt{source\_bus} from the generated topology on module 3. This method uses the graph structure to enforce phase consistency while sampling power values based on the discrete hop zones. In the end, this module generates $N$ (defined by the user) samples for every parameter, where each sample represents a distinct, statistically valid realization of the grid, enabling uncertainty quantification.
\subsection{Module 5: Saving into Standard Formats}
Once the synthetic data samples are generated (Module 4), they must be converted into standard formats compatible with power flow solvers. This module provides a streamlined interface to export the synthetic networks to two widely used open-source platforms: \textit{Pandapower} \cite{pandapower_ref_2018} and the \textit{Open Distribution System Simulator} (OpenDSS) \cite{opendss_ref}. The conversion process is abstracted into a single function call, \texttt{save\_synthetic\_network()}. This function accepts the base topological network (generated in Module 3), the paths to the synthetic data samples (generated in Module 4), and the desired output format string (either \texttt{pandapower} or \texttt{opendss}).
\subsection{User Workflow Example}
To demonstrate the practical application of the framework, Listing \ref{lst:workflow} illustrates the end-to-end Python code for generating a synthetic grid from scratch. In this example, the pre-trained models are employed to generate samples for a topology defined by a 1.5 km radius around São Paulo, Brazil.
\begin{lstlisting}[language=Python, caption={Python script for generating a synthetic grid from scratch using \texttt{bayesgrid}.}, label={lst:workflow}]
import bayesgrid as bg

# Step 1: Initialize Bayesian Models (Module 1) 
bpm = bg.BayesianPowerModel()
bim = bg.BayesianImpedanceModel()
bfm = bg.BayesianFrequencyModel()
bdm = bg.BayesianDurationModel()

# Step 2: Generate Topology (Module 3) 
graph, hop_zone_list, source_bus = bg.create_osm_pandapower_network(
    query_type="address",
    query_address="Sao Paulo, Brazil",
    dist=1500
)

# Step 3: Synthesis via Transfer Learning (Module 4) 
bpm.generate_consistent_data(graph, hop_zone_list, source_bus, n=1000)
bim.generate_consistent_data(graph, hop_zone_list, source_bus, n=1000)
bfm.generate_consistent_data(graph, hop_zone_list, source_bus, n=1000)
bdm.generate_consistent_data(graph, hop_zone_list, source_bus, n=1000)

# Step 4: Export to Simulation Formats (Module 5) 
bg.save_synthetic_network(
    base_net=graph, 
    format="opendss", 
)
\end{lstlisting}
\vspace{-1em}
\section{Case Studies}
We demonstrate the functionality of \texttt{bayesgrid} tool for three distinct case studies. The first focuses on a detailed analysis of a generated synthetic networks for a specific location, validating the real-life coherence of the electrical and reliability parameters produced by the tool. The second case study tests the framework's scalability and spatial sensitivity by generating multiple grids across diverse geographical locations worldwide. The third one includes a probabilistic hosting capacity analysis using the generated samples. For a more detailed validation including learning performance, computational analysis and transfer learning ability, the reader may refer to our previous works \cite{caetano2026bayesian,Caetano2025_powertech,Caetano2024}.
\subsection{Case Study I: Visual Validation of a Synthetic Grid}
\begin{figure*}[htbp]
    \centering
    \begin{subfigure}[b]{0.29\linewidth}
        \centering
        \includegraphics[width=\linewidth]{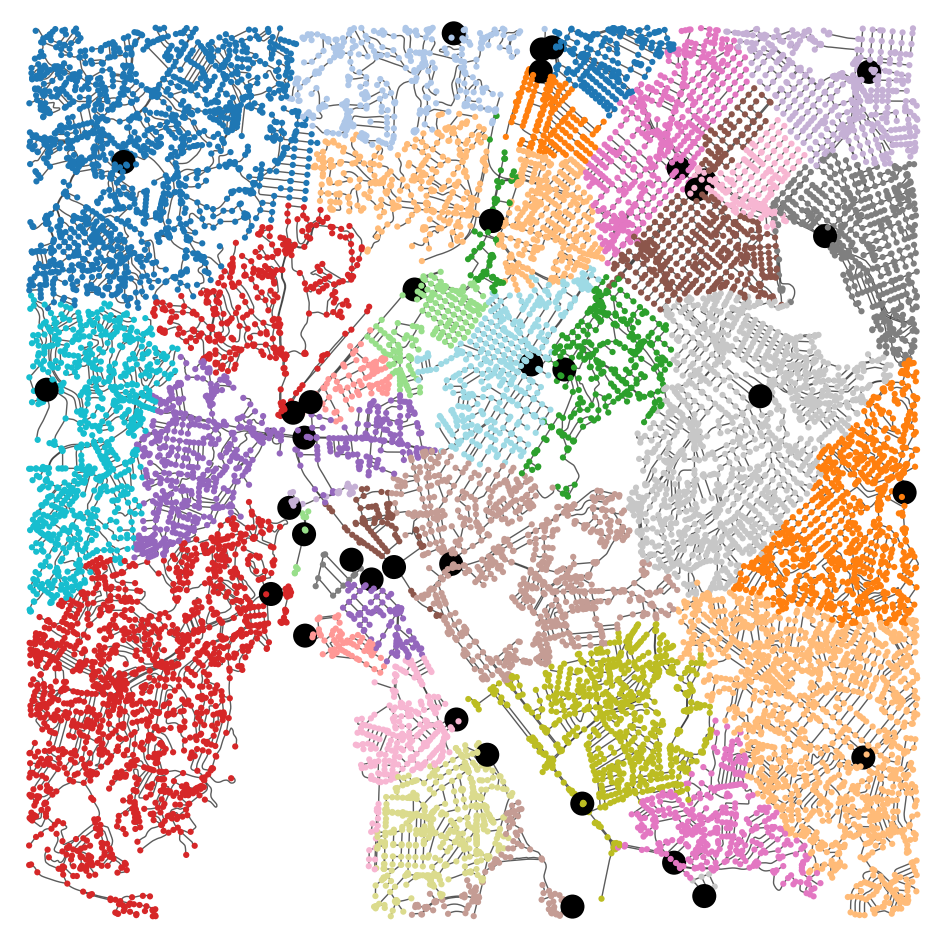}
        \caption{Estimated grid topology}
        \label{fig:topo_a}
    \end{subfigure}
    \hfill 
    \begin{subfigure}[b]{0.32\linewidth}
        \centering
        \includegraphics[width=\linewidth]{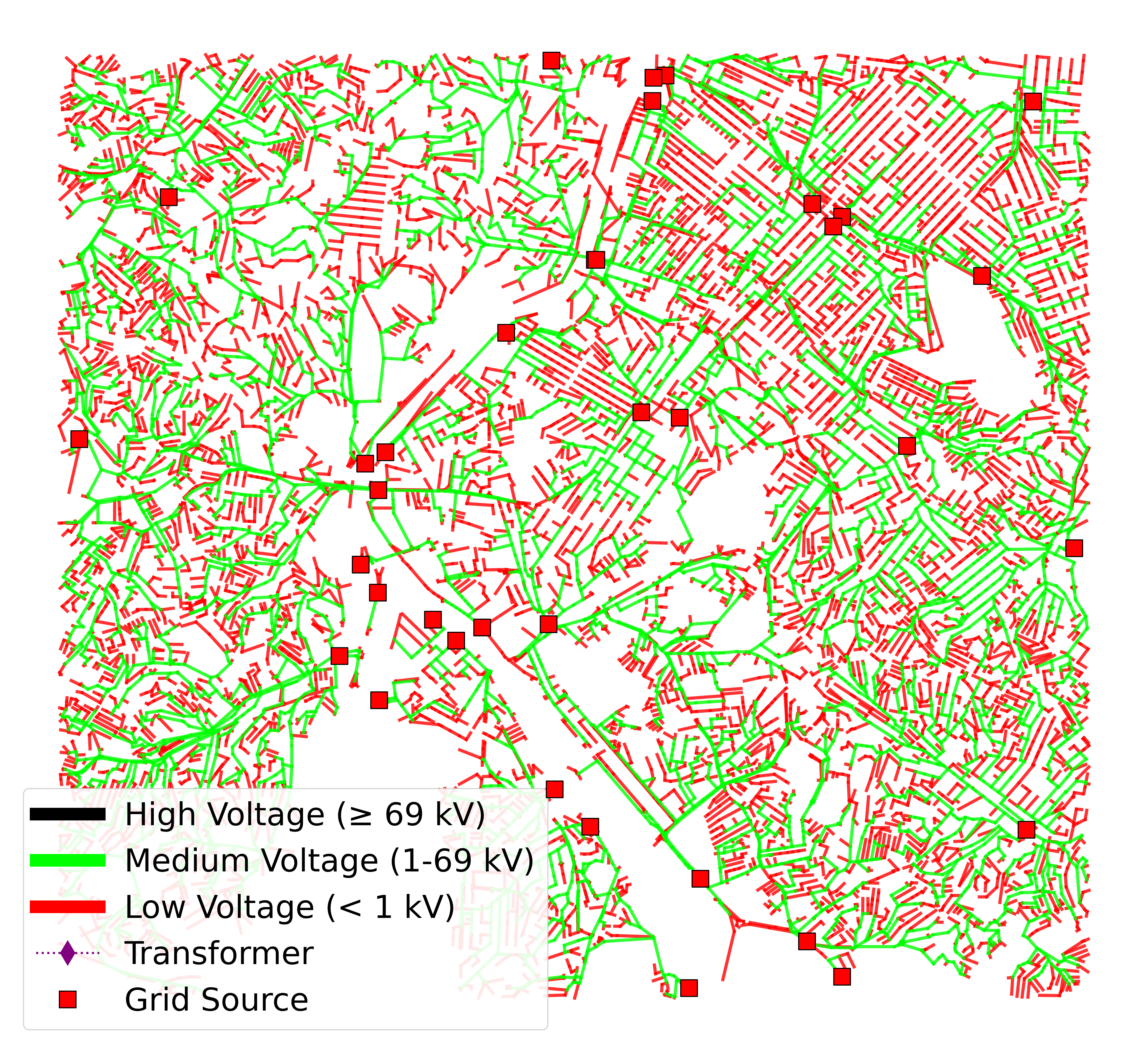}
        \caption{Distribution level}
        \label{fig:topo_b}
    \end{subfigure}
    \hfill 
    \begin{subfigure}[b]{0.32\linewidth}
        \centering
        \includegraphics[width=\linewidth]{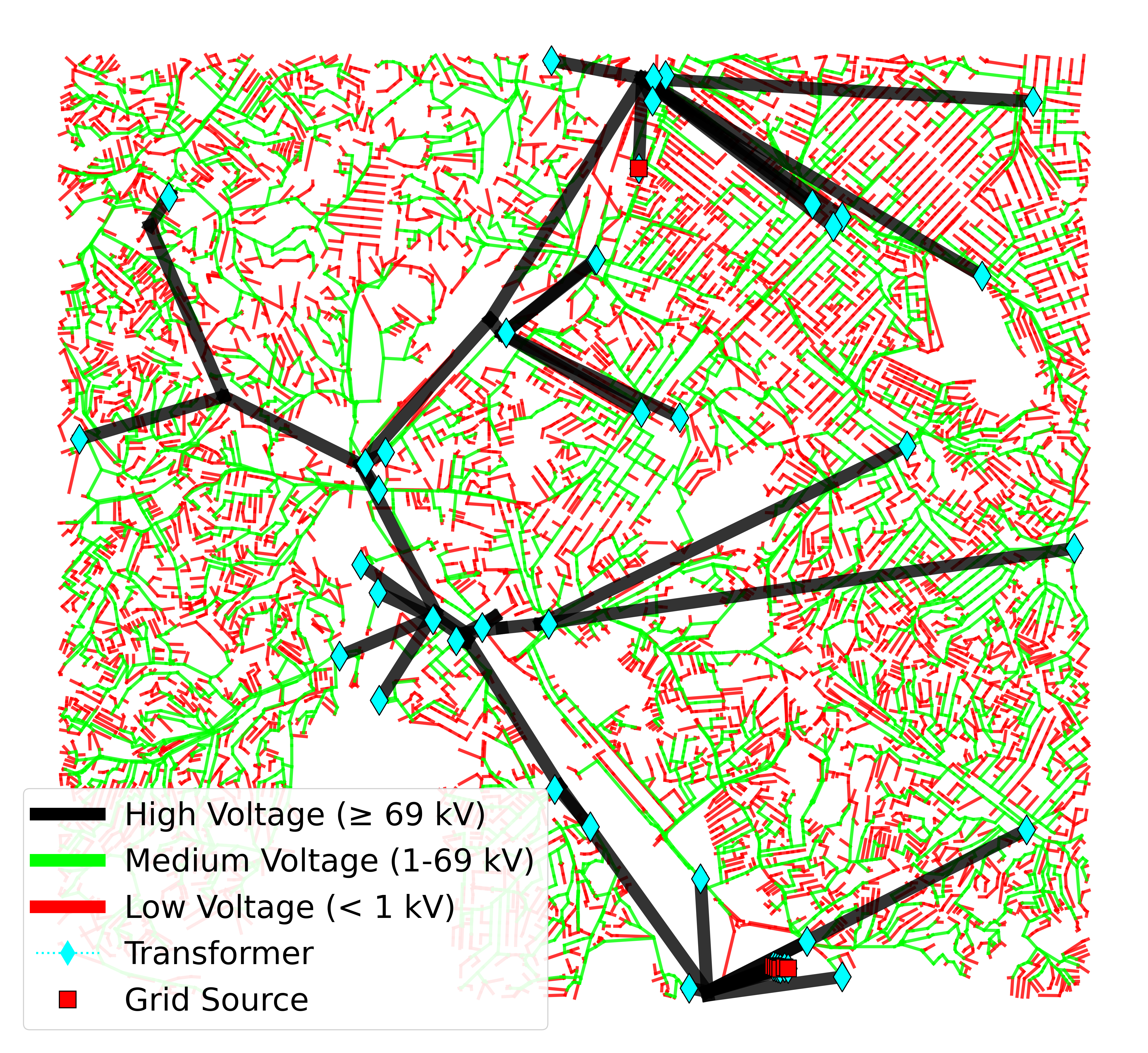}
        \caption{Transmission level}
        \label{fig:topo_c}
    \end{subfigure}
    \caption{Topology of the synthetic network generated from OpenStreetMap data for a 6 km radius around coordinates $(-23.649, -46.702)$. (a) The estimated grid topology, where black dots denote primary substations and distinct node colors define the service area supplied by each substation. (b) Distribution level transformers and voltage levels. (c) Transmission level integrated with high-voltage network.}
    \label{fig:topology}
    \vspace{-1em}
\end{figure*}

This study utilizes the pre-trained version of \texttt{bayesgrid}, leveraging the priors learned from the BDGD dataset. The topology for the synthetic network was generated using the `Point + Radius' query method centered at  $\texttt{query\_point} = (-23.649, -46.702)$ with a radius of  $\texttt{dist} = 7km$. The resulting grid's topology is shown in Figure \ref{fig:topology}. The remaining results of this Case Study will focus on a single feeder of the distribution grid shown in Figure \ref{fig:topology}, for visualization purposes.

\begin{figure}[!htbp]
    \vspace{-1.5em}
    \centering
    \includegraphics[width=\linewidth]{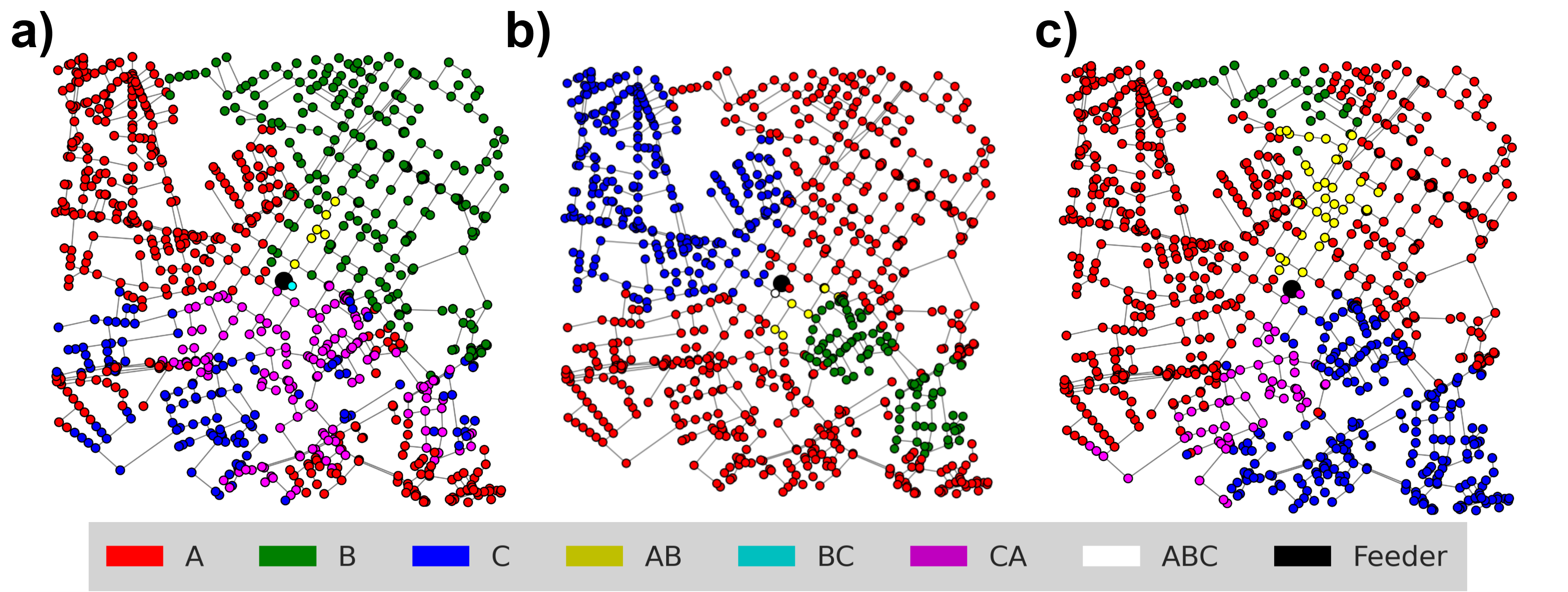}
    \caption{Three distinct samples of phase allocation, all using the starting topology shown in Figure \ref{fig:topology}.}
    \label{fig:phase_samples_together}
    \vspace{-1em}
\end{figure}

A critical requirement for realistic distribution modeling is \textit{phase consistency}. Figure \ref{fig:phase_samples_together} presents three distinct samples drawn from the Bayesian posterior distribution for the same topology. The results demonstrate the stochastic nature of the tool, as each sample exhibits a unique phase configuration. All samples adhere to physical constraints, where downstream branches maintain valid sub-phase configurations relative to their upstream parent nodes. This variation highlights the model's ability to generate an ensemble of valid grid realizations rather than a single deterministic output.

Table \ref{tab:phase_allocation_stats} further quantifies this variability by aggregating the phase allocation probabilities across all generated samples. The 97\% Highest Density Interval (HDI) indicates the uncertainty range for each configuration, confirming that while certain phases (like single-phase A) are dominant, there is significant stochastic variation across the ensemble.

\begin{table}[hbt!]
\centering
\caption{Phase allocation statistics across the generated synthetic samples, showing the Mean probability and the 94\% Highest Density Interval (HDI).}
\label{tab:phase_allocation_stats}
\begin{tabular*}{\columnwidth}{@{\extracolsep{\fill}} cccc}
\toprule
\textbf{Phase Configuration} & \textbf{Mean (\%)} & \textbf{HDI 3\%} & \textbf{HDI 97\%} \\
\midrule
A   & 39.93 & 0.00 & 98.32 \\
B   & 29.64 & 0.00 & 86.96 \\
C   & 26.46 & 0.00 & 79.92 \\
AB  & 0.82  & 0.00 & 3.89  \\
BC  & 2.37  & 0.00 & 10.83 \\
AC  & 0.65  & 0.00 & 4.42  \\
ABC & 0.12  & 0.11 & 0.21  \\
\bottomrule
\end{tabular*}
\end{table}

\begin{figure}[htbp!]
    \centering
    \includegraphics[width=\linewidth]{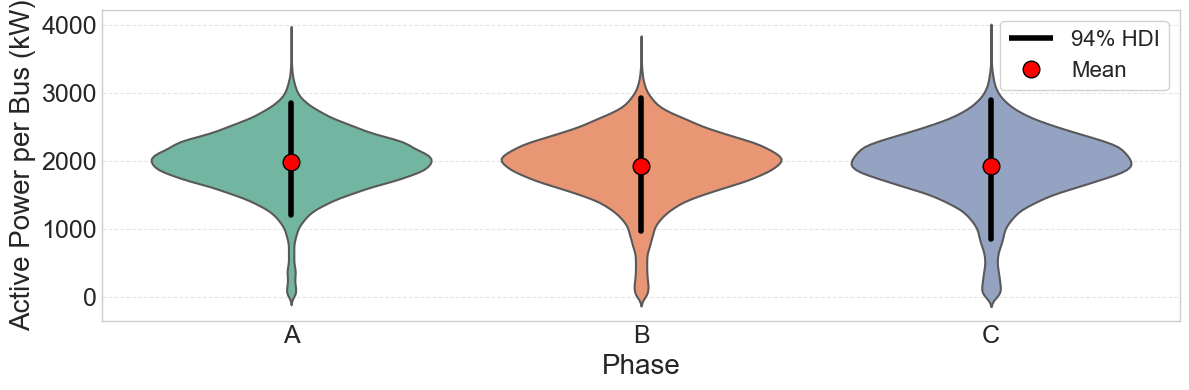}
    \caption{Posterior distribution of the total active power demand across all synthetic samples.}
    \label{fig:power_histogram_sp}
\end{figure}

Figure \ref{fig:power_histogram_sp} presents the distribution of total active power demand across all generated samples for the generated networks. The spread of the distribution reflects the inherent variability in load profiles captured by the BHM, providing a probabilistic range for the system's expected load rather than a single point estimate.

\begin{figure}[htbp!]
    \vspace{-1em}
    \centering
    \includegraphics[width=\linewidth]{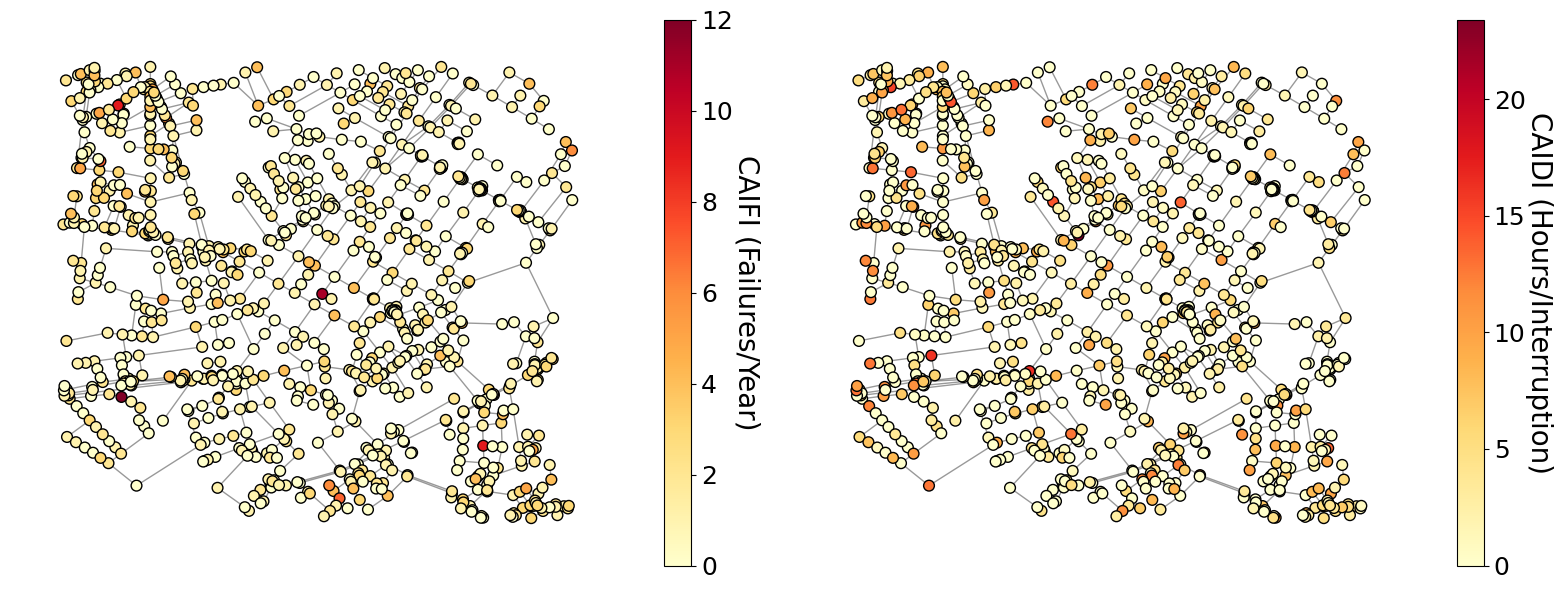}
    \caption{Spatial distribution of generated reliability metrics for a single sample of the synthetic network, using the starting topology shown in Figure \ref{fig:topology}.}
    \label{fig:caifi_and_caidi}
\end{figure}

Figure \ref{fig:caifi_and_caidi} visualizes the generated reliability indices for a single sample. The left panel displays the Customer Average Interruption Frequency Index (CAIFI), modeled via the Negative Binomial distribution, while the right panel shows the Customer Average Interruption Duration Index (CAIDI), modeled via the Hurdle-Weibull distribution. The spatial distribution of these metrics reflects the learned dependency on topological distance, capturing the stochastic nature of failure events across the network.

\begin{figure}[htbp!]
    \centering
    \begin{subfigure}[b]{0.45\textwidth}
        \centering
        \includegraphics[width=\linewidth]{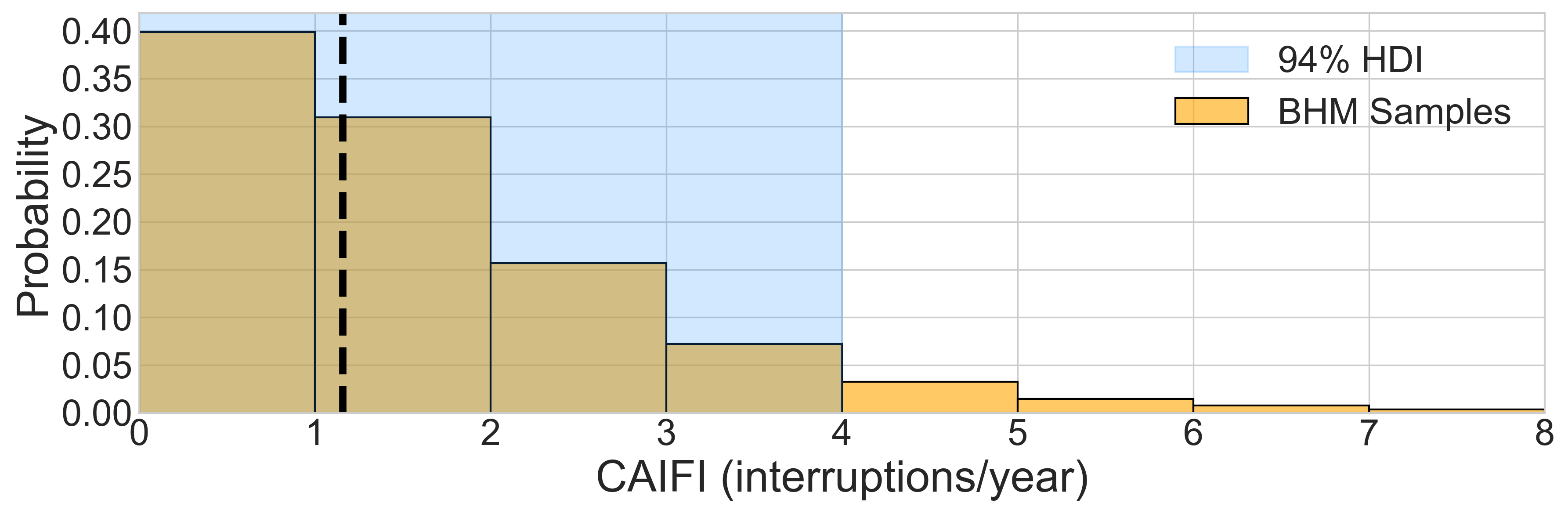}
        \caption{CAIFI Distribution.}
        \label{fig:caifi_histogram_sp}
    \end{subfigure}
    \hfill
    \begin{subfigure}[b]{0.45\textwidth}
        \centering
        \includegraphics[width=\linewidth]{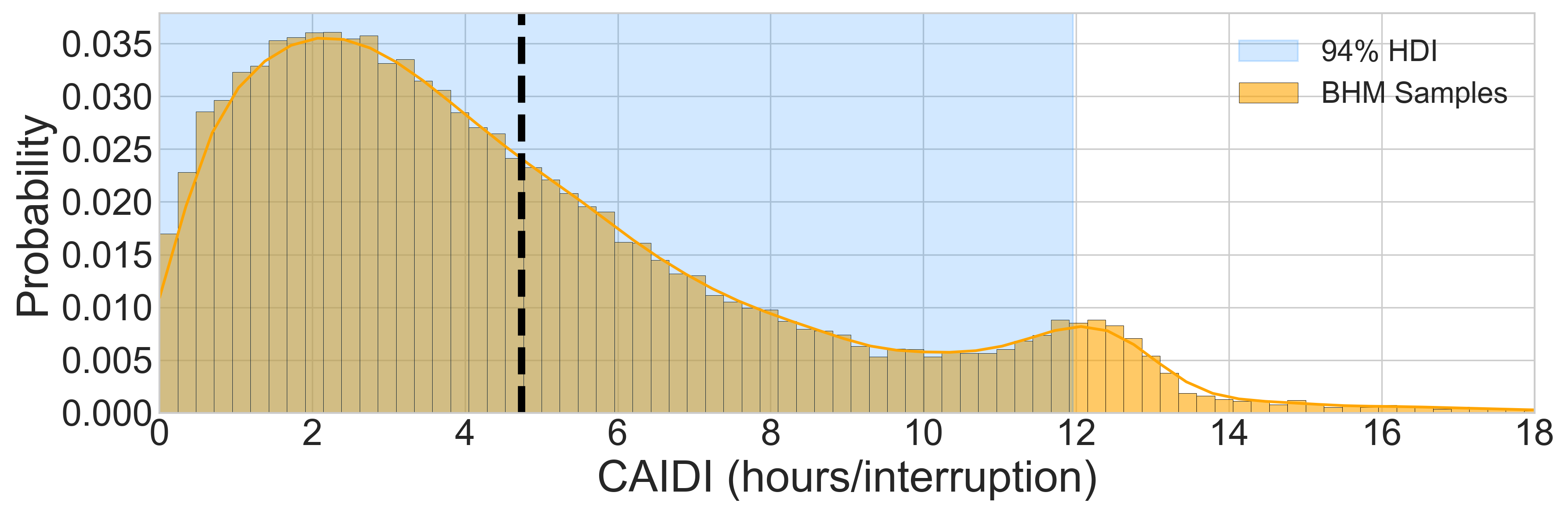}
        \caption{CAIDI Distribution.}
        \label{fig:caidi_histogram_sp}
    \end{subfigure}
    \caption{Posterior distributions of reliability indices aggregated over all synthetic samples. (a) Distribution of average system frequency (CAIFI). (b) Distribution of average system duration (CAIDI).}
    \label{fig:caifi_and_caidi_histogram_sp}
\end{figure}

Figure \ref{fig:caifi_and_caidi_histogram_sp} shows the distributions of these metrics across all generated samples. The histograms for CAIFI (Fig. \ref{fig:caifi_histogram_sp}) and CAIDI (Fig. \ref{fig:caidi_histogram_sp}) reveal the full range of probable reliability scenarios for this grid topology. This probabilistic output is particularly valuable for risk assessment, allowing operators to evaluate the likelihood of extreme reliability events beyond average performance.

\begin{figure}[htbp!]
    \centering
    \includegraphics[width=\linewidth]{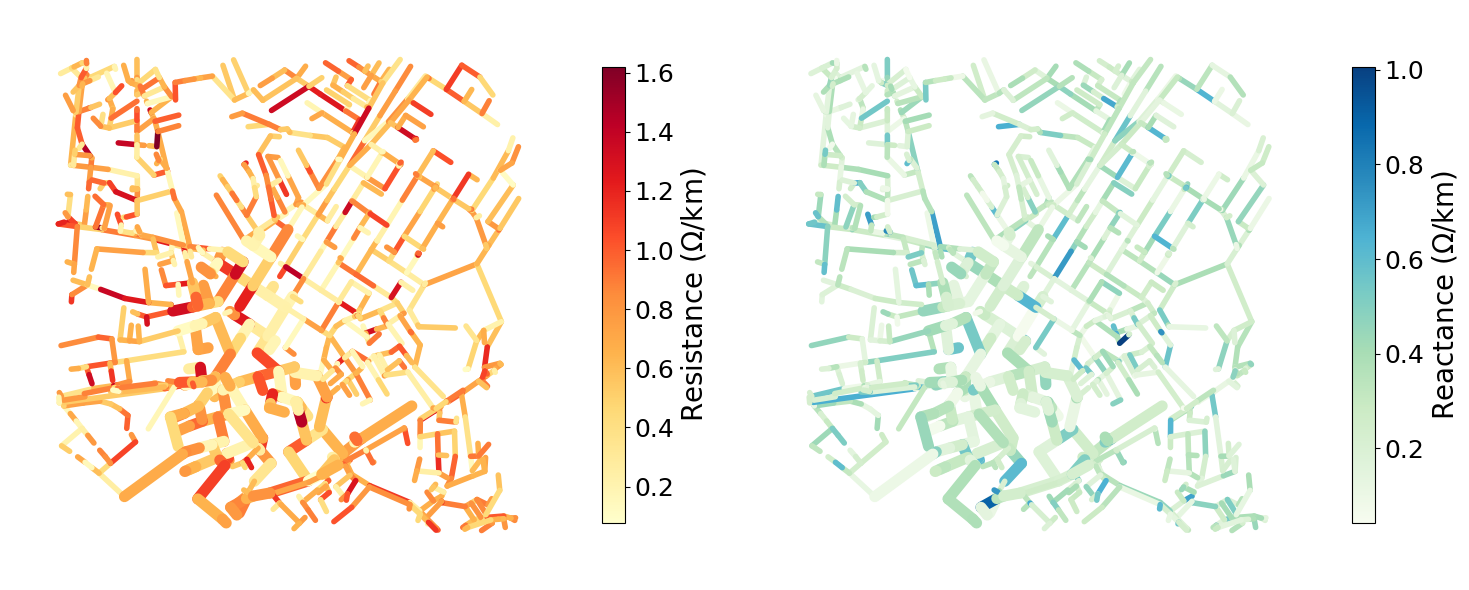}
    \caption{Generated positive-sequence line parameters for a single sample of the synthetic grid, for the topology shown in Figure \ref{fig:topology}. Line thickness corresponds to the number of phases (1, 2, or 3).}
    \label{fig:r_and_x_sao_paulo}
\end{figure}

Finally, the generated line impedance parameters for a single sample are shown in Figure \ref{fig:r_and_x_sao_paulo}. The positive-sequence resistance ($R_1$) and reactance ($X_1$) are plotted, with line thickness proportional to the number of phases. The results exhibit consistency in the $R/X$ ratio, which remains approximately 1.5, a typical value for distribution systems.

\begin{figure}[hbt!]
    \centering
    \begin{subfigure}[b]{0.45\textwidth}
        \centering
        \includegraphics[width=\linewidth]{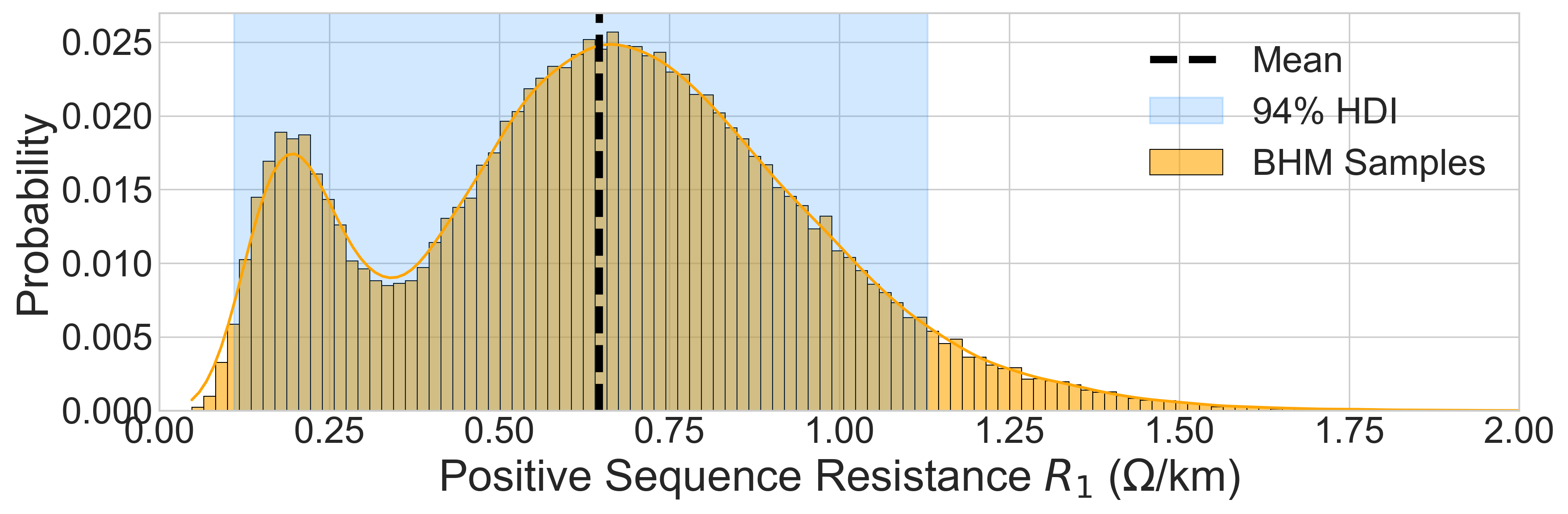}
        \caption{Resistance ($R_1$) Distribution.}
        \label{fig:r_histogram_sp}
    \end{subfigure}
    \hfill
    \begin{subfigure}[b]{0.45\textwidth}
        \centering
        \includegraphics[width=\linewidth]{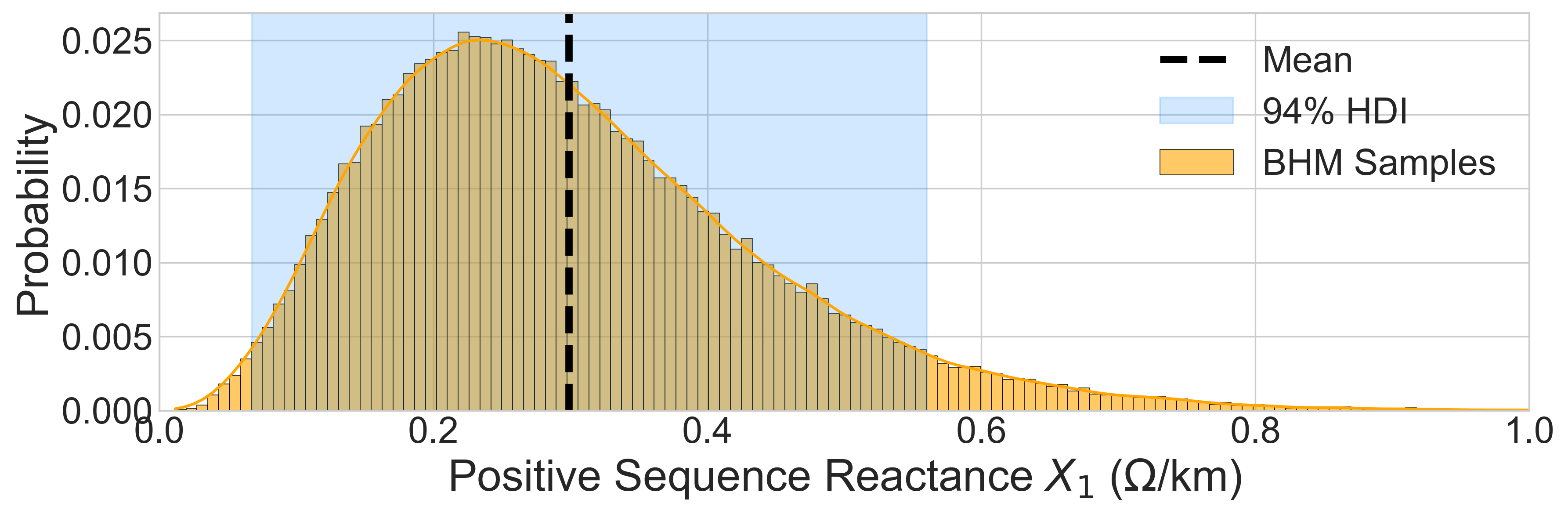}
        \caption{Reactance ($X_1$) Distribution.}
        \label{fig:x_histogram_sp}
    \end{subfigure}
    \caption{Posterior distributions of the mean line impedance parameters across all synthetic samples. (a) Distribution of mean Resistance. (b) Distribution of mean Reactance.}
    \label{fig:r_and_x_histogram_sp}
\end{figure}

Similar to the previous metrics, Figure \ref{fig:r_and_x_histogram_sp} displays the histograms for the mean resistance and reactance across all samples. These distributions confirm that while the topology is fixed, the electrical characteristics of the lines vary stochastically, allowing for robust Monte Carlo simulations that account for parameter uncertainty in impedance modeling.

\begin{figure}[htbp!]
    \centering
    \includegraphics[width=\linewidth]{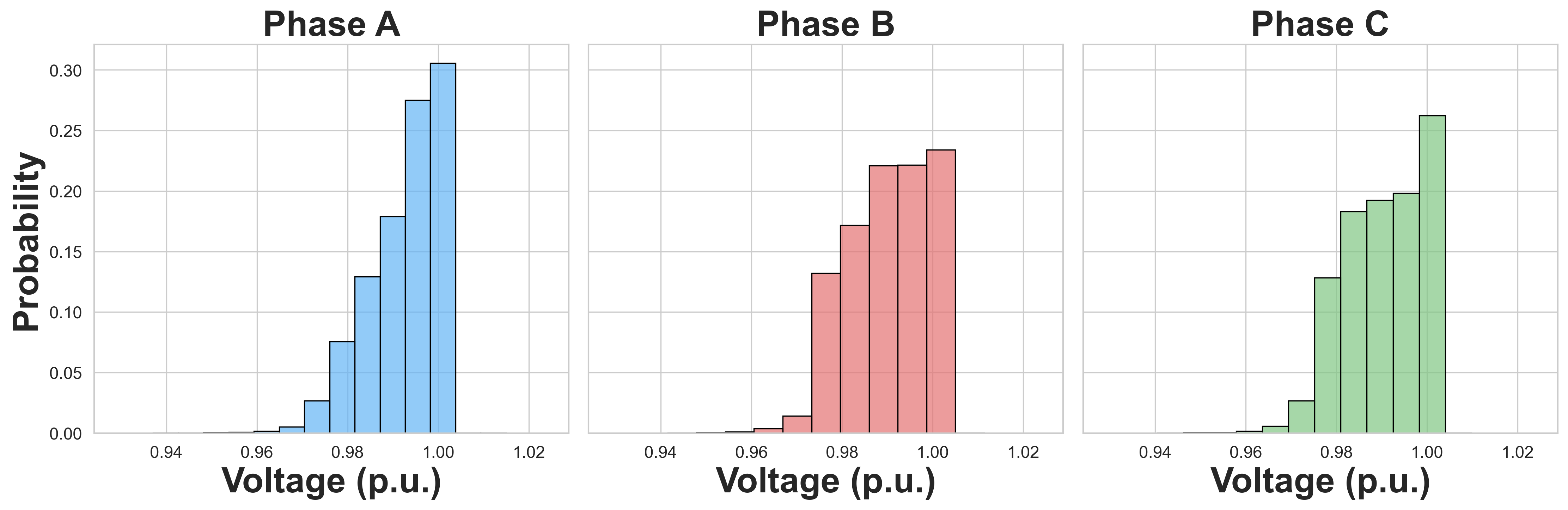}
    \caption{Probability distribution of nodal voltages (p.u.) for Phases A, B, and C across all synthetic grids generated worldwide.}
    \label{fig:voltage_level_worldwide}
\end{figure}

Beyond the statistical coherence of the individual parameters, it is essential to verify that their combination yields a functioning electrical network. To assess this, we performed unbalanced three-phase power flow simulations on every generated sample using OpenDSS. The \texttt{bayesgrid} framework demonstrated robust performance, achieving a 100\% convergence rate for the power flow algorithms across all synthetic realizations. Regarding power quality, Figure~\ref{fig:voltage_level_worldwide} presents the aggregated nodal voltage profiles from all generated cities worldwide. Standard regulations require voltage levels to remain between $0.9$ and $1.1$ p.u. \cite{Gupta2021}. As shown, the simulated steady-state voltages successfully adhere to this viable operational range across all three phases.
\subsection{Case Study II: Generalization and Scalability}

\begin{figure*}[htbp]
    \centering
    \begin{subfigure}{0.19\linewidth}
        \centering
        \includegraphics[width=\linewidth]{figures/sao_paulo_voronoi.png}
        \caption{São Paulo}
    \end{subfigure}
    \begin{subfigure}{0.19\linewidth}
        \centering
        \includegraphics[width=\linewidth]{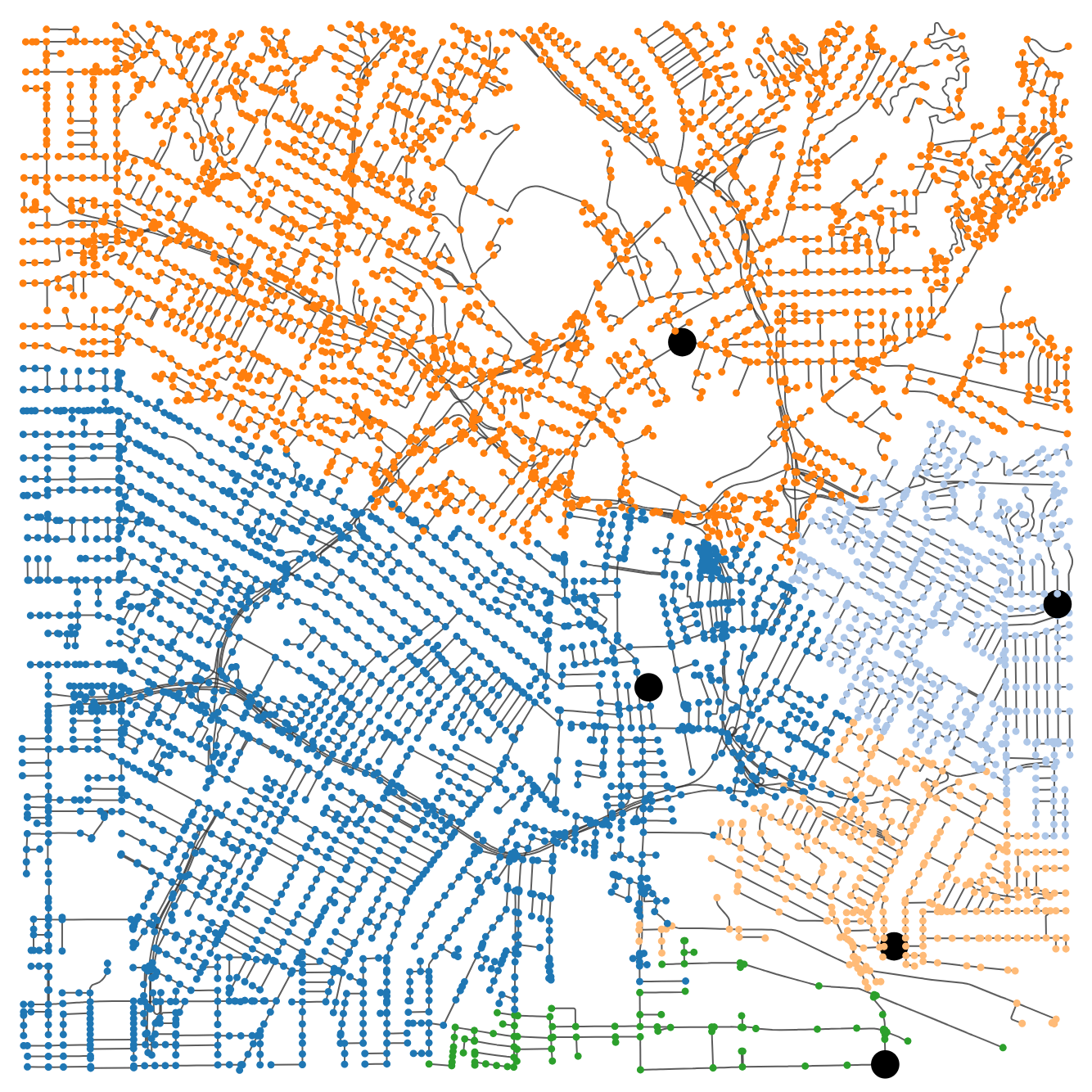}
        \caption{Los Angeles}
    \end{subfigure}
    \begin{subfigure}{0.19\linewidth}
        \centering
        \includegraphics[width=\linewidth]{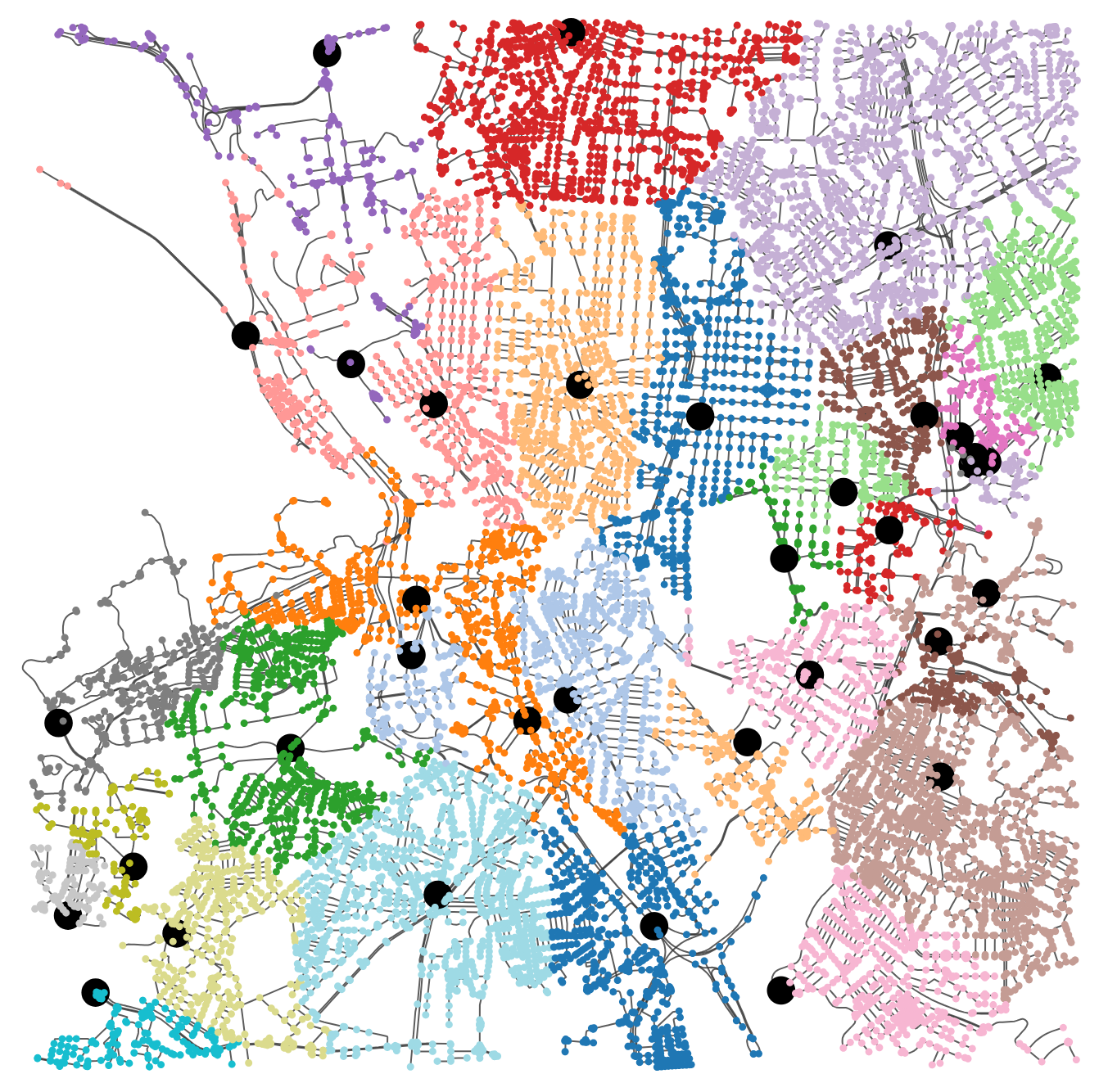}
        \caption{Madrid}
    \end{subfigure}
    \begin{subfigure}{0.19\linewidth}
        \centering
        \includegraphics[width=\linewidth]{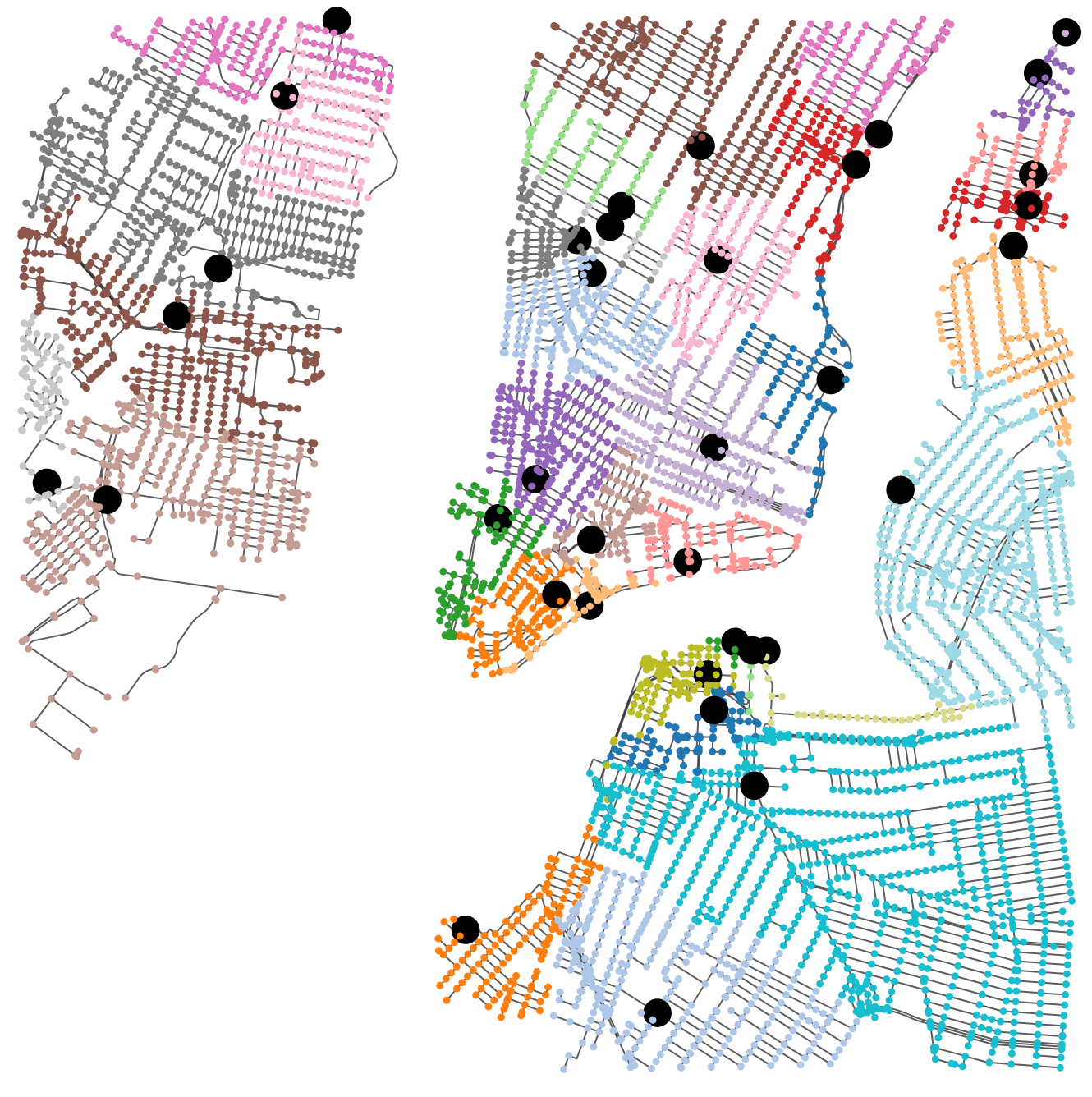}
        \caption{New York}
    \end{subfigure}
    \begin{subfigure}{0.20\linewidth}
        \centering
        \includegraphics[width=\linewidth]{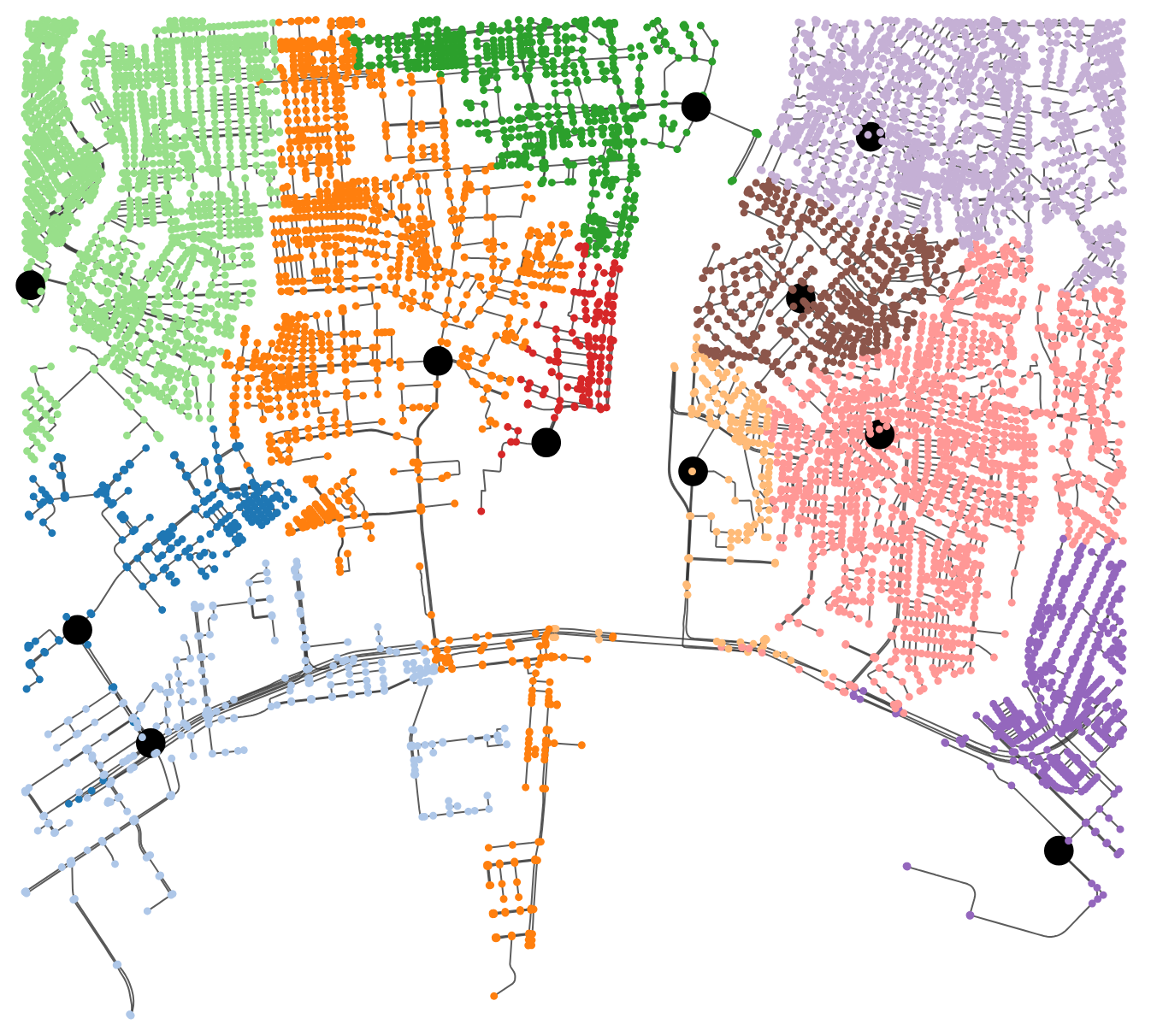}
        \caption{Tokyo}
    \end{subfigure}
    \caption{Synthetic network topologies generated for diverse global cities. Each color represents a distinct service area for each substation.}
    \label{fig:global_cities}
\end{figure*}
To demonstrate the global applicability and computational scalability of the framework, we generated synthetic networks for five major metropolitan areas across different continents: São Paulo (Brazil), Los Angeles (USA), Madrid (Spain), New York (USA), and Tokyo (Japan). This study evaluates the tool's performance in handling diverse grid topologies and large-scale networks.
\begin{table*}[hbt!]
\centering
\caption{Computational performance for synthetic grid generation across five global cities.}
\label{tab:comp_time_global}
\begin{tabular*}{\textwidth}{@{\extracolsep{\fill}} l c c c c c}
\toprule
\textbf{Metric} & \textbf{São Paulo} & \textbf{Los Angeles} & \textbf{Madrid} & \textbf{New York} & \textbf{Tokyo} \\
\midrule
\textbf{Number of Buses} & 11,228 & 7,626 & 14,943 & 6,832 & 10,112 \\
\textbf{Computational time to generate the grid's topology} & 1m 03s & 0m 52s & 1m 26s & 0m43s & 0m57s \\
\midrule
\textbf{Synthetic Network Generation, per sample average (ms)} & & & & & \\
Phase Allocation \& Power Demand & 362  & 252 & 422 & 240 & 354 \\
Interruption Frequency (CAIFI) & 13 & 22 & 24 & 19 & 15 \\
Interruption Duration (CAIDI) & 17 & 23 & 13 & 16 & 19 \\
Line Parameters (R1 \& X1) & 320 & 285 & 343 & 248 & 311 \\
\bottomrule
\end{tabular*}
\vspace{-2em}
\end{table*}
Table \ref{tab:comp_time_global} reports the computation time for generating a single synthetic sample for networks of varying sizes. Even for the largest network (Tokyo, with over 50,000 nodes), the generation of all electrical and reliability parameters takes less than a second per sample. The topology generation step, which involves querying and processing OSM data, scales well with the complexity of the street network but remains practical for large-scale studies, taking less than 2 minutes for city-scale grids.

To verify that the statistically generated parameters yield physically viable networks, we performed power flow simulations on the synthetic grids for all five cities. Across all locations and samples, the power flow algorithms achieved a 100\% convergence rate. Furthermore, 100\% of the generated samples maintained nodal voltages within the standard acceptable range of $[0.9, 1.1]$ p.u. 
\subsection{Case Study III: Probabilistic Hosting Capacity}
To demonstrate the utility of \texttt{bayesgrid} in risk-aware infrastructure planning, we conducted a probabilistic Hosting Capacity (HC) analysis. All power flow simulations were conducted using OpenDSS. To ensure realistic load behaviors, typical load profiles were sourced from the Brazilian National Utility Data (BDGD) \cite{bdgd_ref}, consistent with the dataset used to train the original Bayesian models. Solar irradiance data was obtained from the PVGIS database (European Commission Joint Research Centre) \cite{pvgis}, and we mapped the latitude and longitude of each load in the synthetic grid to the corresponding PVGIS coordinates, generating a mean irradiance curve based on hourly historical data from 2005 to 2023. The probabilistic HC assessment methodology follows the approach detailed in \cite{masic2025hosting, Gupta2021}.

\begin{figure}[htbp!]
    \centering
    \begin{subfigure}[b]{0.45\textwidth}
        \centering
        \includegraphics[width=\linewidth]{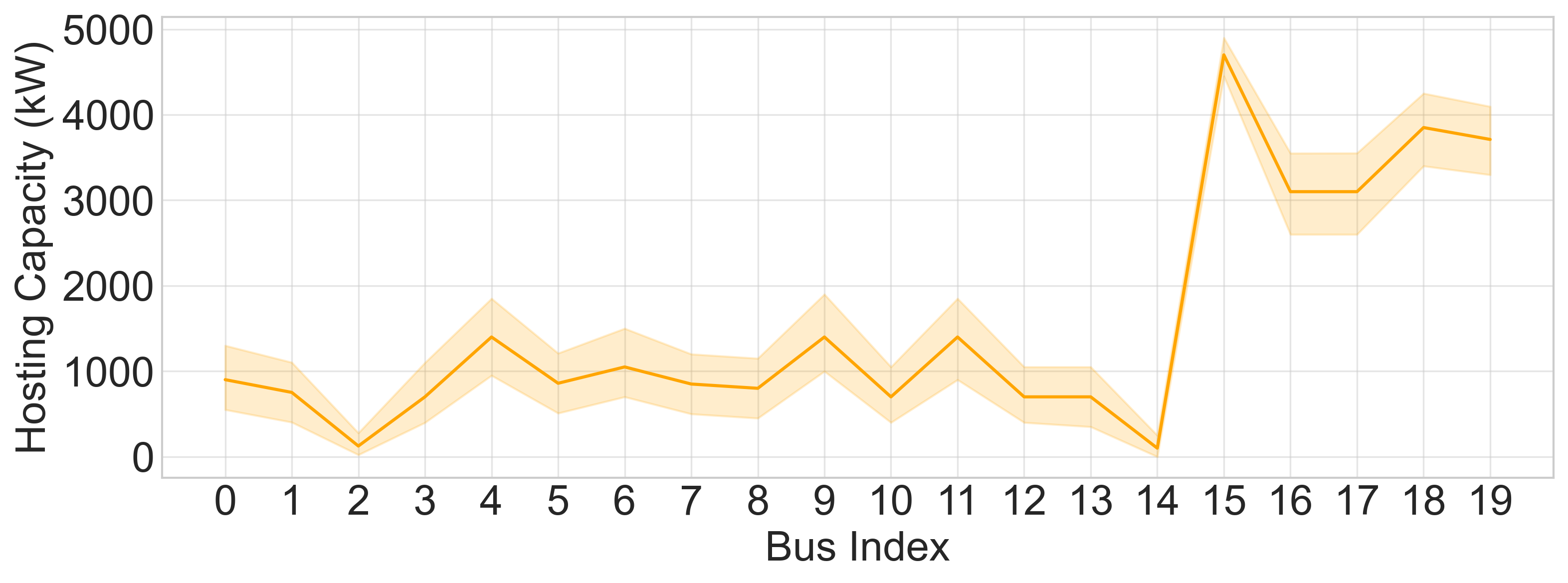}
        \caption{Per-bus hosting capacity uncertainty}
        \label{fig:per_bus_hc}
    \end{subfigure}
    \hfill
    \begin{subfigure}[b]{0.45\textwidth}
        \centering
        \includegraphics[width=\linewidth]{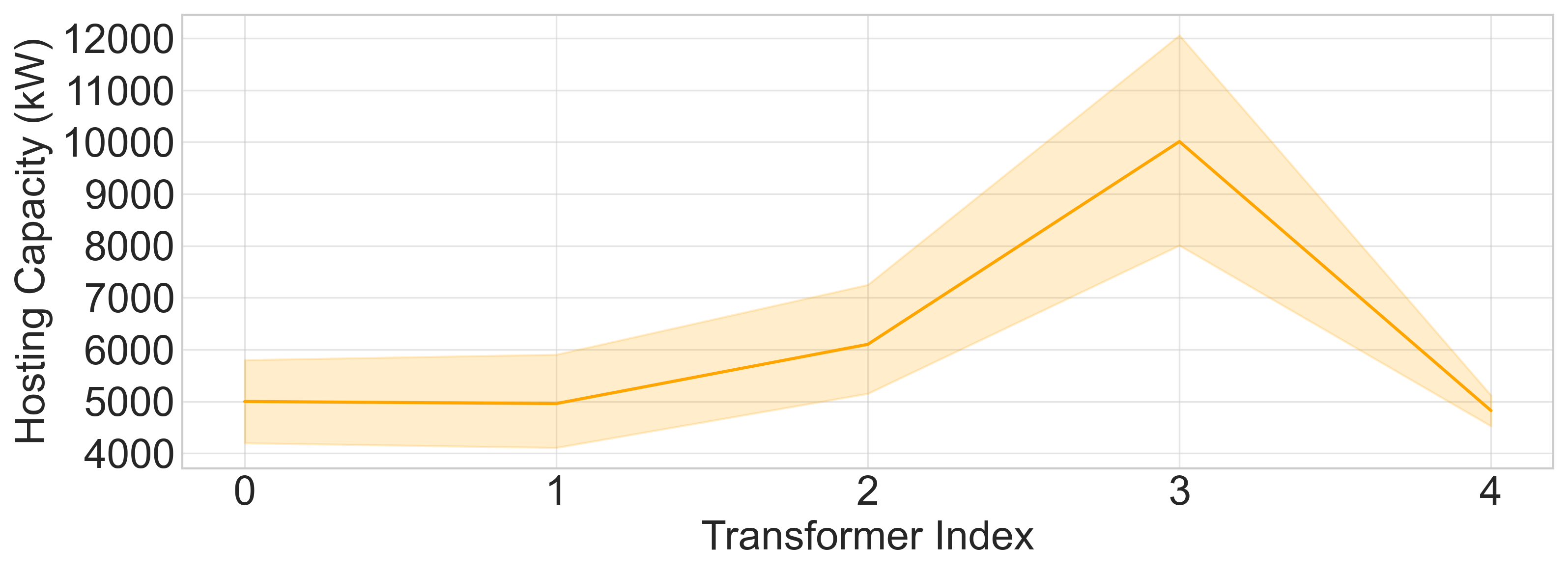}
        \caption{Per-transformer hosting capacity uncertainty}
        \label{fig:per_transformer_hc}
    \end{subfigure}
    \caption{Per-bus and per-transformer hosting capacity analysis for the São Paulo grid, shown in Figure \ref{fig:phase_samples_together}.}
    \label{fig:hosting_capacity_sp}
\end{figure}

The analysis was performed on two distinct topologies to highlight different granularities of uncertainty. The first set of results were conducted on the smaller São Paulo topology shown in Figure \ref{fig:phase_samples_together}, focusing on bus-level and transformer-level HC. Figure \ref{fig:hosting_capacity_sp} illustrates the HC variability across $N=10,000$ synthetic samples. As shown in Figure \ref{fig:per_bus_hc}, the hosting capacity for individual buses is not a static value but a range, represented by the confidence intervals. Similarly, Figure \ref{fig:per_transformer_hc} shows how specific transformers exhibit varying HC across different synthetic samples. This variability allows planners to identify weak points in the grid that may statistically be more prone to higher HC levels.

\begin{figure}[htbp!]
    \vspace{-1em}
    \centering
    \begin{subfigure}[b]{0.45\textwidth}
        \centering
        \includegraphics[width=\linewidth]{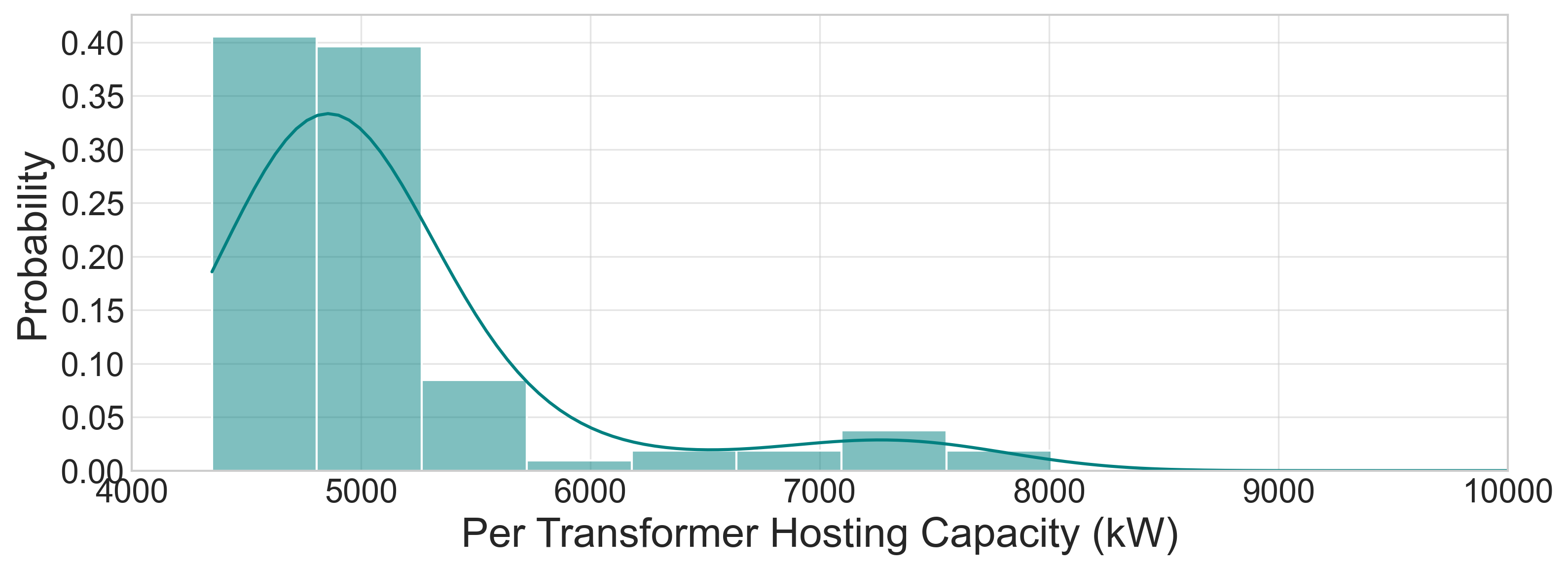}
        \caption{Distribution of Per-Transformer HC}
        \label{fig:los_angeles_per_transformer_hc}
    \end{subfigure}
    \hfill
    \begin{subfigure}[b]{0.45\textwidth}
        \centering
        \includegraphics[width=\linewidth]{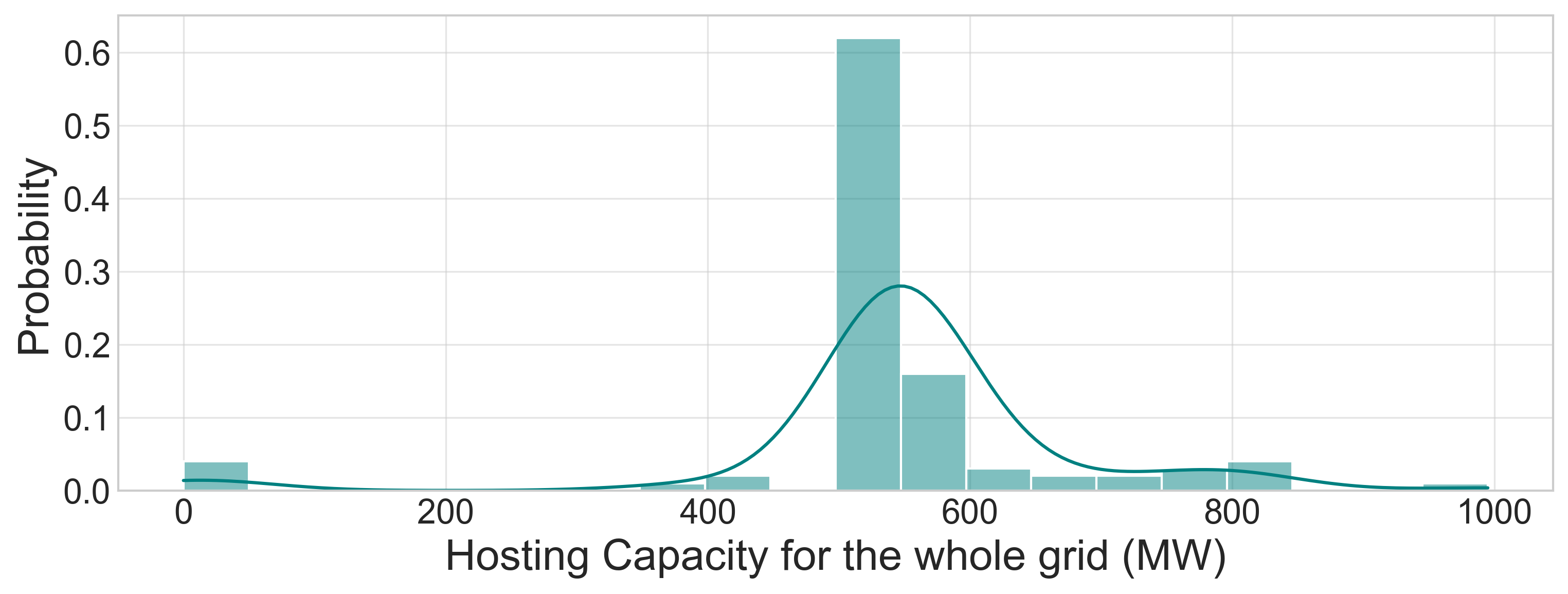}
        \caption{System-Level Total HC Distribution}
        \label{fig:los_angeles_system_hc}
    \end{subfigure}
    \caption{Per-transformer and System-level hosting capacity analysis for the Los Angeles grid, shown in Figure \ref{fig:global_cities}.}
    \label{fig:hosting_capacity_la}
    \vspace{-1em}
\end{figure}

The hosting capacity was also computed for the Los Angeles grid, shown in Figure \ref{fig:global_cities}(b). Given the large number of buses, we present the analysis by aggregating the HC to the transformer and system levels. Figure \ref{fig:los_angeles_per_transformer_hc} presents the probability distribution of mean per-Transformer HC across all samples. Furthermore, Figure \ref{fig:los_angeles_system_hc} aggregates this to the system level, providing a macroscopic view of the grid's total solar installation potential. These distributions highlight the influence of probabilistic modeling across the generated samples, evidenced by the large uncertainty in the observed HC values at both the transformer and system levels.

These results demonstrate that \texttt{bayesgrid} enables the probabilistic study for a given use-case such as hosting capacity analysis. 
By quantifying the uncertainty of hosting capacity, grid operators can make investment decisions based on statistical confidence rather than deterministic assumptions, a capability not possible with traditional deterministic synthetic network approaches.
\section{Conclusion}
\label{sec:conclusion}
The accelerating energy transition and the increasing complexity of distribution systems, driven by the integration of distributed energy resources and evolving demand patterns, have created an urgent need for robust, data-driven modeling tools. However, the research community continues to face a significant bottleneck: the scarcity of detailed and large-scale open-access network data due to security and confidentiality constraints. Moreover, current tools lacks generalization, wide applicability and they are deterministic in nature. To address this gap, this paper developed and introduced \texttt{bayesgrid}, an open-source Python tool designed to generate realistic, synthetic power distribution systems.

Unlike existing synthetic network tools that typically produce a deterministic and static network representation for a given region, \texttt{bayesgrid} distinguishes itself in generating probablistic networks enabled by the developed open-source Python framework that apply Bayesian Hierarchical Models to treat grid parameters as stochastic variables. By integrating topological data from OpenStreetMap with statistical learning, the tool goes beyond simple topology generation to explicitly model complex distribution characteristics often overlooked in the literature, specifically unbalanced three-phase power and reliability indices (CAIFI, CAIDI). The generated distribution network is interfaced with estimated transmission network leveraging the public OSM data. Furthermore, the tool demonstrates direct interoperability with standard simulation platforms, automatically exporting valid, ready-to-run network ensembles for both \textit{Pandapower} and \textit{OpenDSS}.

We demonstrated the tool's scalability and transfer learning capabilities through the generation of synthetic networks across several real metropolitan areas. The results confirmed that the generated networks are electrically feasible, achieving 100\% power flow convergence and valid voltage profiles with negligible computational cost during the sampling phase. Future work will focus on expanding its physical modeling capabilities. Key directions include the probabilistic allocation of active components, such as distributed generation (PV), battery storage systems, and Electric Vehicle Charging Stations.  

\bibliographystyle{IEEEtran}
\bibliography{references.bib}

\vspace{-33pt}

\vfill

\end{document}